\documentclass[
aps,
prd,
superscriptaddress,
nofootinbib,
floatfix]
{revtex4}
\usepackage{graphicx,
longtable,
color,
array,booktabs,
bm}
\usepackage[colorlinks=true,
           linkcolor=blue,
           urlcolor=blue,
           citecolor=blue]{hyperref}

\newcommand{\Ge}{$^{76}$Ge}
\newcommand{\Te}{$^{130}$Te}
\newcommand{\Xe}{$^{136}$Xe}
\newcommand{\Mo}{$^{100}$Mo}
\begin{document}
\title{Interplay between non-interfering neutrino exchange mechanisms \\ 
and nuclear matrix elements in $0\nu\beta\beta$ decay
}
%
\author{        	Eligio~Lisi}
\email{				eligio.lisi@ba.infn.it}
\affiliation{   	Istituto Nazionale di Fisica Nucleare, Sezione di Bari, 
               		Via Orabona 4, 70126 Bari, Italy}
\author{        	Antonio~Marrone}
\email{				antonio.marrone@ba.infn.it}
\affiliation{   	Dipartimento Interateneo di Fisica ``Michelangelo Merlin,'' 
               		Via Amendola 173, 70126 Bari, Italy}%
\affiliation{   	Istituto Nazionale di Fisica Nucleare, Sezione di Bari, 
               		Via Orabona 4, 70126 Bari, Italy}
\author{        	Newton~Nath}
\email{				newton.nath@ba.infn.it}
\affiliation{   	Istituto Nazionale di Fisica Nucleare, Sezione di Bari, 
               		Via Orabona 4, 70126 Bari, Italy}
%

\begin{abstract}
We revisit the phenomenology of neutrinoless double beta ($0\nu\beta\beta$) decay  mediated by non-interfering exchange of light and heavy Majorana neutrinos, in the context of current and prospective ton-scale experimental searches, as well as of recent calculations of nuclear matrix elements (NME) in different nuclear models. 
We derive joint upper bounds on the light and heavy contributions to $0\nu\beta\beta$ decay, for different sets of NME, through separate and combined data coming from the following experiments (and isotopes): KamLAND-Zen and EXO (Xe), GERDA and MAJORANA (Ge) 
and CUORE (Te). We further consider three proposed projects that could provide, within current bounds, possible $0\nu\beta\beta$ decay signals at $>\!3\sigma$ level with an exposure of 10 ton years:  nEXO (Xe), LEGEND (Ge) and CUPID (Mo). Separate and combined (Xe, Ge, Mo) signals are studied for different representative cases and NME sets, and the conditions leading to (non)degenerate light and heavy neutrino mechanisms  are discussed. In particular, the role of heavy-to-light NME ratios in different isotopes is highlighted through appropriate graphical representations. By using different sets of ``true'' and ``test'' NME as a proxy for nuclear uncertainties, it is shown that the relative contributions of light and heavy neutrino exchange to $0\nu\beta\beta$ signals may be significantly biased in some cases. Implications for theoretical models connecting light and heavy Majorana neutrino masses are also briefly illustrated. These results provide further motivations to improve NME calculations, so as to better exploit the physics potential of future multi-isotope $0\nu\beta\beta$ searches at the ton scale.
\end{abstract}
\maketitle

\section{Introduction}
\label{Sec:Intro}

The search for neutrinoless double beta decay ($0\nu\beta\beta$) in various $(Z, A)$ isotopes,
\begin{equation}
(Z,\,A) \to (Z+2, \,A) + 2e^-\ ,
\label{0nubb}
\end{equation}
violating the lepton number by two units, represents a major research program in particle and nuclear physics \cite{Agostini:2022zub,Adams:2022jwx,Acharya:2023swl}. 
The observation of such rare process
would prove that neutrinos are Majorana particles \cite{Schechter:1981bd}, independently of the particle physics mechanism(s) leading to the decay, as reviewed, e.g., in \cite{Agostini:2022zub,Rodejohann:2011mu,Vergados:2012xy}. 

Assuming the simplest mechanism involving the exchange of the three known light neutrinos, the decay half-life $T_i$ for the isotope $i=(Z, A)$ reads
\begin{equation}
\label{Tlight}
(T_i)^{-1} =S_i = G_i\, M^2_{\nu,i}\, m^2_\nu\ ,
\end{equation}
where $G_i$ is the phase-space factor \cite{Deppisch:2020ztt,Stoica:2019ajg}, $M_{\nu,i}$ is the nuclear matrix element (NME) 
\cite{Engel:2016xgb} and $m_\nu$ is the so-called effective Majorana mass for light $\nu$,
\begin{equation}
\label{mlight}
m_\nu = \left|\sum_{k=1}^3 U^2_{ek}m_k\right|\ ,
\end{equation}
where $U_{ek}$ is the mixing matrix element relating  $\nu_e$ to the light state $\nu_k$ with mass $m_k$. We follow a previously adopted notation 
\cite{Capozzi:2021fjo,Lisi:2022nka} 
by introducing the signal strength $S_i=1/T_i$, and absorbing in $G_i$ terms as  $1/m^{2}_e$ and $g^4_A$, where $g_A=1.276$ 
\cite{Markisch:2018ndu} is the bare value of the axial-vector coupling.
We can make contact with the alternative notation of \cite{Agostini:2022zub}, where $1/T = G_{01} g_A^4M^2_{0\nu}m^2_{\beta\beta}/m^2_e$,
by identifying $m_\nu=m_{\beta\beta}$, as well as $G=G_{01}g_A^4/m^2_e$ and $M_\nu=M_{0\nu}$ for each isotope $i$. 
As in \cite{Agostini:2022zub}, we take   
the NME values $M_{\nu,i}$ as positive real numbers, referred to the bare value of $g_A$. 
 

%
The pursuit of understanding the 
origin of possible $0\nu\beta\beta$ decay processes 
has led to an exploration of alternative scenarios, that could either replace or coexist with the 
exchange of light Majorana neutrinos $\nu_k$
\cite{Agostini:2022zub,Rodejohann:2011mu,Vergados:2012xy}. 
Of particular importance is the exchange of heavy Majorana neutrinos ($N_h$), that are a crucial ingredient
of the celebrated seesaw mechanism \cite{Minkowski:1977sc,Gell-Mann:1979vob,Yanagida:1979as,Glashow:1979nm,Mohapatra:1979ia,Schechter:1980gr}, a
fundamental framework aimed at addressing the smallness of neutrino masses. In this framework, heavy Majorana neutrinos
may have a potential impact on various leptonic processes, including $0\nu\beta\beta$ decay. 

The significance of heavy Majorana neutrinos in the context of $0\nu\beta\beta$ decay has been recognized in various studies \cite{Halprin:1976mr,Halprin:1983ez,Bamert:1994qh,Atre:2009rg,Bilenky:2011tr}. The presence of both heavy  ($N_h$) and light  ($\nu_k$) Majorana neutrinos introduces an interplay of contributions that can enrich the phenomenology of $0\nu\beta\beta$ decay. In particular, within the seesaw framework, the heavy and light neutrinos sectors 
are connected, and $0\nu\beta\beta$ data can constrain them jointly, at least in principle.
Furthermore, low-scale left-right (LR) gauge models, a class of LR-symmetric models, have emerged as interesting candidates to 
generate $0\nu\beta\beta$ decay \cite{Tello:2010am,Barry:2012ga,Chakrabortty:2012mh,Das:2012ii,Deppisch:2017vne,Fukuyama:2022naj,Patra:2023ltl}. 
Such LR models have gained attention not only for their potential to explain the observed neutrino masses and 
mixing but also for their broader implications for unifying electroweak and strong interactions \cite{Pati:1974yy,Mohapatra:1974gc,Senjanovic:1975rk}. These models introduce new particles and interactions, e.g., right-handed (RH) currents, that could manifest in $0\nu\beta\beta$ decay as well. 

A notable avenue in this quest involves considering the roles played by both $\nu_k$ and $N_h$ \cite{Halprin:1983ez}, which find significance in 
many scenarios beyond the Standard Model (SM) \cite{Mohapatra:2005wg}. In this context, 
$0\nu\beta\beta$ decay amplitudes due to both light and heavy neutrino exchange may appear, and their possible interference effects need to be
considered. 
In this work we focus on the case of non-interfering (incoherent) contributions, that arise in phenomenologically 
interesting scenarios \cite{Halprin:1983ez} and particularly in LR-symmetric models, where heavy neutrinos such as $N_h$ are connected to RH 
currents through massive $W_R$ bosons ($m_{W_R}\gg m_{W}\simeq 80.4$ GeV) \cite{Mohapatra:1974gc,Senjanovic:1975rk}.
For a recent investigation of the dynamics of interference suppression in LR symmetric models, see~\cite{Ahmed:2019vum}. Hereafter, we assume 
the case of non-interfering contributions of light and heavy neutrinos to $0\nu\beta\beta$ decay; comments on interfering contributions are given below.

For non-interfering (incoherent) exchange of $\nu_k$ and $N_h$, Eq.~(\ref{Tlight}) is generalized 
with the same phase space \cite{Simkovic:1999re} as:
\begin{equation}
\label{Tlightheavy}
(T_i)^{-1} =S_i = G_i\, \left(M^2_{\nu,i}\, m^2_\nu\ + M^2_{N,i}\, m^2_N \right)\ ,
\end{equation}
where the $M_{N,i}$ represent the NME for heavy Majorana neutrino exchange, while $m_N$ is another effective Majorana mass parameter for heavy $N_h$ that, in LR models and in our notation, typically takes the form (see, e.g., \cite{Rodejohann:2011mu})
\begin{equation}
\label{mheavy}
m_N = \frac{m_{W}^4}{m_{W_R}^4}\left|\sum_{h} V^2_{eh}\frac{m_p\,m_e}{M_h}\right|\ ,
\end{equation}
$M_h$ being the mass of the heavy $N_h$, and $V_{eh}$ the associated mixing matrix element. Hereafter, we shall take
Eq.~(\ref{Tlightheavy}) as our working hypothesis for a phenomenological analysis of current and prospective $0\nu\beta\beta$ decay data.
   
In principle, if the NME for both light and heavy neutrino exchange were accurately known in two different
isotopes $i$ and $j$, two precise experimental signals $S_i$ and $S_j$ would be sufficient to determine the 
two unknown  parameters $m_\nu$ and $m_N$ via 
the coupled equations 
\begin{equation}
\label{System}
\left[
\begin{array}{c}
S_i G_i^{-1}\\ 
S_j G_j^{-1} 
\end{array}
\right] = 
\left[
\begin{array}{cc}
M^2_{\nu,i} & M^2_{N,i} \\ 
M^2_{\nu,j} & M^2_{N,j} 
\end{array}
\right]
\left[
\begin{array}{c}
m^2_\nu \\ 
m^2_N 
\end{array}
\right] \ ,
\end{equation}
provided that their determinant is nonzero, namely, that the heavy-to-light NME ratios are isotopically different \cite{Deppisch:2006hb}, 
\begin{equation}
\label{Ratios}
\frac{M_{N,i}}{M_{\nu,i}} \neq \frac{M_{N,j}}{M_{\nu,j}}\ .
\end{equation}
An additional signal in a third isotope $k$ (with another, different NME ratio) would then act as a consistency check. Conversely,
competing $0\nu\beta\beta$ mechanisms with very similar NME ratios would be largely degenerate. 

Note that algebraic conditions equivalent to Eq.~(\ref{Ratios})
are also required to disentangle and check interfering mechanisms, where the signal strength is of the form $S_i = G_i |M_{\nu,i}\, m_\nu + M_{N,i}\, m_N |^2$, namely, a coherent sum of amplitudes.
In this case, however, additional complications arise due to the emergence of unknown relative phases among the different amplitudes, that may lead to either constructive 
or destructive interference, see e.g.\ \cite{Pascoli:2013fiz}. Various phenomenological studies, from early ones with up to four different mechanisms \cite{Faessler:2011qw,Faessler:2011rv} 
to more recent investigation with two (standard plus exotic) amplitudes \cite{Agostini:2022bjh}, show that cancelations play an important role in enlarging the 
allowed space of parameters ($m_\nu$ and $M_N$ in our case), especially if the NME ratios lead to degeneracies. In this sense, the scenario with
non-interfering (incoherent) $0\nu\beta\beta$ mechanisms considered herein is relatively simple with respect to the case of coherent mechanisms.
 
In practice, even this simple non-interfering scenario for a multi-isotope determination of $m_\nu$ and $m_N$
is hindered by several problems: ($i$) the NME are currently affected by large uncertainties, not necessarily (all) reduced by taking ratios; ($ii$) the ratios $M_{N,i}/M_{\nu,i}$ happen to be quite similar in various isotopes, at least in some nuclear models; ($iii$) available $0\nu\beta\beta$ data are compatible with null signals while, in perspective, even positive signals may be affected by large statistical uncertainties; ($iv$) multi-isotope signals may lead to consistency checks (if compatible) or to unphysical solutions (if incompatible), depending in part on the assumed NME and their ratios. These and other related issues have been addressed with a variety of approaches and results in a vast literature, with emphasis on different aspects. A largely incomplete list includes  studies of the algebraic NME conditions leading to (non)degenerate mechanisms \cite{Faessler:2011qw,Faessler:2011rv,Simkovic:2010ka} or to (un)physical solutions \cite{Meroni:2012qf}, of general features of light vs heavy NME calculations \cite{Graf:2022lhj,Cirigliano:2018yza,Menendez:2017fdf},
of multi-isotope NME consistency checks \cite{Bilenky:2002ga,Bilenky:2004um}, of 
available or prospective decay rates \cite{Gehman:2007qg,Lisi:2015yma,Neacsu:2021mwo} and of additional spectral data that may help to break degeneracies \cite{Horoi:2015gdv}, just to name a few topics. 
It may also be noticed that the apparent similarity of NME ratios ${M_{N,i}}/{M_{\nu,i}}$ in different isotopes 
can be regarded, on the one hand, as a disadvantage, leading to an effective degeneracy of light and heavy mechanisms; 
and on the other hand as an advantage, 
leading to an isotope-independent generalization of Eq.~(\ref{Tlightheavy}) that interpolates between the light and heavy $\nu$ mass scales \cite{Babic:2018ikc}, covering
the possible regime of intermediate masses (not considered in this work) at the Fermi momentum scale of $O(200)$ MeV. 

Despite the difficulties in unraveling the above issues, $0\nu\beta\beta$ decays mediated by light and heavy neutrinos continue to attract interest, both theoretically and experimentally.
Several theoretical frameworks that establish a connection between the light and heavy sectors offer the possibility of testing relationships between parameters such as $m_\nu$ and $m_N$ in $0\nu\beta\beta$ processes. Additionally, these models link processes that violate lepton number at both low-energy and high-energy scales (e.g., at colliders~\cite{Kersten:2007vk}). For comprehensive overviews, refer to the reviews in \cite{Deppisch:2015qwa, Bolton:2019pcu}. Early work can be found in \cite{Tello:2010am}, and more recent studies are presented in \cite{Patra:2023ltl}, among many others.
Concerning the NME, theoretical calculations for light and heavy neutrino exchange have been performed for a variety
of candidate isotopes and nuclear models, although with still large uncertainties (as reviewed later). 
A general consensus is emerging about a well-defined roadmap to improve and stabilize the NME calculations 
\cite{Cirigliano:2022oqy,Cirigliano:2022rmf}
by benchmarking the models (possibly based on ab initio techniques) with as many nuclear data as possible, e.g., by exploiting NME correlations 
with a variety of observables \cite{Santopinto:2018nyt,Romeo:2021zrn,Ejiri:2022zdg,Yao:2022usd,Jokiniemi:2022ayc,Jokiniemi:2023bes,Jiao:2023poq}. 
A recent implementation of this program for $^{136}$Xe suggests an encouraging reduction of the associated NME uncertainties (formally below
20\% at $1\sigma)$ \cite{Horoi:2023uah}, although outstanding problems remain, such as the role of $g_A$ quenching \cite{Ejiri:2019ezh,Suhonen:2017krv}
or the assessment of the relative sign and size of some short-range contributions to the decay rate
\cite{Cirigliano:2018hja,Jokiniemi:2021qqv,Pompa:2023jxc}, with a possible different impact for light and heavy neutrino exchange.  

From the experimental viewpoint, half-life constraints $T_i>10^{25}$~y have been placed by five experiments in three isotopes: 
KamLAND-Zen \cite{KamLAND-Zen:2022tow} and EXO \cite{Anton:2019wmi} ($^{136}$Xe), GERDA \cite{Agostini:2020xta}
and MAJORANA \cite{Majorana:2022udl} ($^{76}$Ge), and CUORE \cite{CUORE:2021mvw} ($^{130}$Te). The next important goal 
will be to reach sensitivities $T_i\sim 10^{28}$~y (and possibly first signals at $>3\sigma$ level) 
by using detector masses of about 1~ton operating on a decadal timescale;
multi-isotope searches will remain crucial to cross-check the results and to test the
underlying mechanism(s) \cite{Agostini:2022zub,Adams:2022jwx,Acharya:2023swl}. A prospective international program  
\cite{Acharya:2023swl} envisages three ton-scale projects using different isotopes, 
such as  nEXO \cite{nEXO:2021ujk} ($^{136}$Xe),  LEGEND-1000 \cite{LEGEND:2021bnm} ($^{76}$Ge) and CUPID-1T \cite{CUPID:2022wpt} ($^{100}$Mo).  
Connecting this low-energy program with high-energy searches for heavy neutral leptons will provide complementary tests of 
neutrino  physics beyond the standard model   \cite{Abdullahi:2022jlv}.

\textcolor{black}{In this evolving context, we think it appropriate 
to revisit in detail several aspects of the phenomenology of $0\nu\beta\beta$ decay with non-interfering light and heavy neutrino exchange mechanisms.
Our work includes a comprehensive set of recent NME from various nuclear models, a state-of-the-art 
analysis of both current $0\nu\beta\beta$ data and prospective decay signals in ton-scale detectors, and a discussion
of the effects induced by the spread of NME (and of their heavy-to-light $\nu$ 
ratios), with emphasis on separate and joint bounds in the parameter space ($m^2_\nu,\,m^2_N$), 
where phenomenological results and theoretical predictions can be usefully illustrated.} In particular,  
we discuss the 
NME values and their ratios as obtained in different nuclear models for the (Xe, Ge, Te, Mo) isotopes in the last decade
 (Sec.~\ref{Sec:NME}). We perform 
an up-to-date  statistical analysis of 
the most constraining data from current (Xe, Ge, Te) experiments for all the NME sets (Sec.~\ref{Sec:Current}). We also
analyze prospective $0\nu\beta\beta$ signals 
observable  at $>3\sigma$ level in ton-scale (Xe, Ge, Mo) projects, for representative NME sets and values of $m_\nu$ and $m_N$ (Sec.~\ref{Sec:Future}). 
Effects of nuclear model uncertainties are studied by swapping ``true'' and ``test'' sets of NME values. 
Cases leading to (non)degenerate light-heavy neutrino mechanisms and to 
(un)biased or (un)physical $m_\nu$ and $m_N$ parameters are discussed. 
We present illustrative tests of a theoretical model connecting $m_\nu$ and $m_N$ (Sec.~\ref{Sec:Model}), and finally
summarize our work (Sec.~\ref{Sec:End}).
In our analysis, the impact of different NME ratios [as in Eq.~(\ref{Ratios})] and the interplay of different bounds
are highlighted through appropriate
graphical representations in terms of squared effective Majorana masses.  Our findings provide additional motivations 
to search for $0\nu\beta\beta$ decay in different isotopes, and to improve the NME calculations for different decay mechanisms.

\section{Sets of nuclear matrix elements}
\label{Sec:NME}

In this work we consider fifteen sets of NME calculated in the last decade,
for both light and heavy neutrino exchange in ($^{136}$Xe, $^{76}$Ge, $^{130}$Te), using different theoretical approaches and their variants:  
the nuclear shell model (SM) \cite{Menendez:2017fdf,Horoi:2015tkc}, the quasi-particle random phase approximation (QRPA)
\cite{Fang:2018tui,Hyvarinen:2015bda,Faessler:2014kka}, the energy-density functional theory (EDF) \cite{Song:2017ktj}, 
and the interacting boson model (IBM) \cite{Deppisch:2020ztt,Barea:2015kwa}. Among these NME sets, eight include 
also calculations for $^{136}$Mo in the QRPA, EDF and IBM models.%
\footnote{The SM approach was recently applied to $^{100}$Mo \protect\cite{Coraggio:2022vgy}, 
but only for the case of light Majorana neutrino exchange.}
 Table~\ref{Tab:NME}
reports the adopted numerical values of the NME for light
and heavy neutrino exchange ($M_\nu$ and $M_N$, respectively, assuming  $g_A=1.276$),
as well as their ratios $M_N/M_\nu$.

\begin{table}[t!]
\centering
\resizebox{.88\textwidth}{!}{\begin{minipage}{\textwidth}
\caption{\label{Tab:NME} 
List of fifteen sets of nuclear matrix elements (NME) for $0\nu\beta\beta$ decay mediated by light neutrinos ($M_{\nu}$) 
 or heavy neutrinos ($M_N$), together
 with the $M_N/M_{\nu}$ ratio. The NMEs are  computed in four isotopes within four different models (SM, QRPA, EDF, and IMB), and refer to
 the bare value $g_A=1.276$.
}
\begin{ruledtabular}
\begin{tabular}{c | ccc | ccc | ccc | ccc| c | c}
NME  & \multicolumn{3}{c |}{\Xe\ } &  \multicolumn{3}{c |}{\Ge\ }  &  \multicolumn{3}{c |}{\Te\ }   &  \multicolumn{3}{c |}{\Mo\ } & Ref. & Model \\ 
set   & $M_{\nu}$  & $M_N$ & $M_N/M_{\nu}$ &  $M_{\nu}$  & $M_N$ & $M_N/M_{\nu}$ & $M_{\nu}$  & $M_N$ & $M_N/M_{\nu}$ &  $M_{\nu}$  & $M_N$ & $M_N/M_{\nu}$ &  &  \\[1mm]
\hline
1 & 2.28  & 116   & 50.87 & 2.89 & 130   & 44.98 & 2.76 & 146   & 52.89    &&& & \cite{Menendez:2017fdf}  &     \\
2 & 2.45  & 167   & 68.16 & 3.07 & 188   & 61.24 & 2.96 & 210   & 70.95    &&& & \cite{Menendez:2017fdf}  &     \\
3 & 1.63  & 98.8  & 60.61 & 3.37 & 126   & 37.38 & 1.79 & 94.5  & 52.79    &&& & \cite{Horoi:2015tkc}     &  SM \\
4 & 1.76  & 143   & 81.25 & 3.57 & 202   & 56.58 & 1.93 & 136   & 70.46    &&& & \cite{Horoi:2015tkc}     &     \\
5 & 2.19  & 114.9 & 52.46 & 2.81 & 132.7 & 47.22 & 2.65 & 144.2 & 54.42    &&& &  \cite{Horoi:2015tkc}    &     \\
%
\hline
%
6  & 1.11 & 66.9  & 60.27  & 3.12 & 187.3 & 60.03 & 2.90 & 191.4 & 66.00 &      &&& \cite{Fang:2018tui}     &     \\
7  & 1.18 & 90.5  & 76.69  & 3.40 & 293.7 & 86.38 & 3.22 & 303.5 & 94.25 &      &&& \cite{Fang:2018tui}     &    \\
8  & 2.91 & 186.3 & 64.02  & 5.26 & 401.3 & 76.29 & 4.00 & 338.3 & 84.57 & 3.9  & 350.8 & 89.95& \cite{Hyvarinen:2015bda}&     \\
9  & 2.75 & 160   & 58.18  & 5.44 & 265   & 48.71 & 4.18 & 240   & 57.42 & 4.79 & 260 & 54.28 &\cite{Vergados:2012xy} & QRPA    \\
10 & 3.36 & 172   & 51.19  & 5.82 & 412   & 70.79 & 4.70 & 385   & 81.91 & 5.15 & 404 & 78.45  & \cite{Vergados:2012xy}     \\
11 & 2.18 & 152   & 69.72  & 5.16 & 287   & 55.62 & 3.89 & 264   & 67.86 & 5.40 & 342 & 63.33 & \cite{Faessler:2014kka}\\
12 & 2.46 & 228   & 92.68  & 5.56 & 433   & 77.87 & 4.37 & 400   & 91.53 & 5.85 & 508 & 86.84 & \cite{Faessler:2014kka} \\
\hline
%
13 & 4.24 & 166.3  & 39.22 & 6.04 & 209.1  & 34.62 & 4.89  & 193.8  & 39.63 & 6.48 & 232.6  & 35.89 &\cite{Song:2017ktj}& EDF   \\
\hline
14 & 3.25 & 97.91  & 30.13 & 5.14 & 157.4  & 30.63 &  3.96 & 124.9  & 31.54 & 3.84 & 115.8  & 30.16  &\cite{Barea:2015kwa} &     \\
15 & 3.40 & 99.17  & 29.17 & 6.34 & 181.6  & 28.65 & 4.15  & 126.8  & 30.56 & 5.07 & 104.1  & 20.54  &\cite{Deppisch:2020ztt} & IBM     \\
\end{tabular}
\end{ruledtabular}
\end{minipage}}
\end{table}

Given the crucial role of isotopically different NME ratios $M_{N,i}/M_{\nu,i}$ to avoid the degeneracy of the two mechanisms [Eq.~(\ref{Ratios})],
it is useful to show such ratios for pairs  of different 
isotopes $(i,\,j)$ relevant in the analysis of current constraints ($i,\,j=~$Xe, Ge, Te) and of prospective signals 
($i,\,j=~$Xe, Ge, Mo).

\begin{figure}[t!]
\begin{minipage}[c]{0.91\textwidth}
\includegraphics[width=0.91\textwidth]{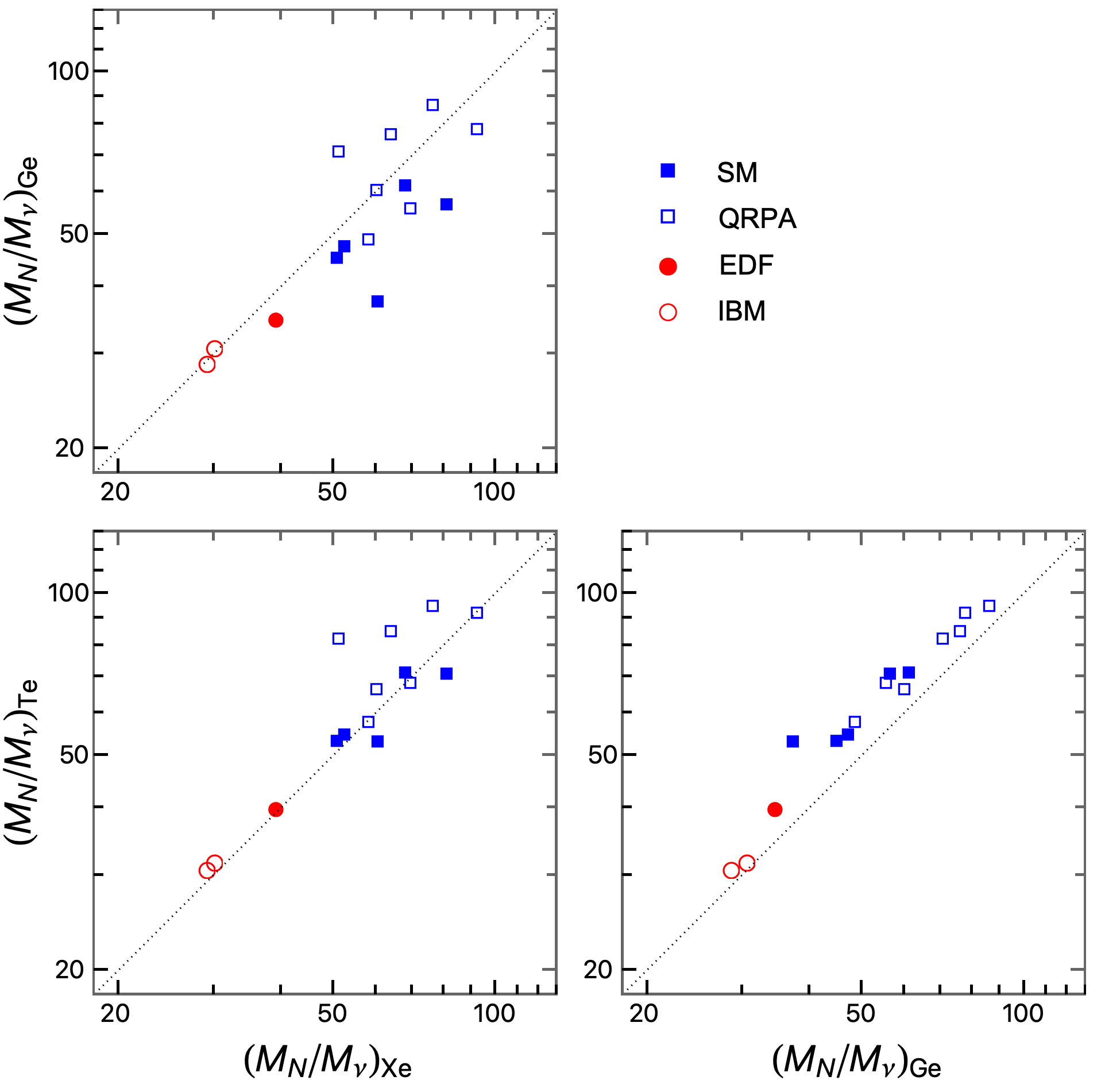}
\vspace*{-2mm}
\caption{\label{Fig_01}
\footnotesize Scatter plot of $M_{N,i}/M_{\nu,i}$ ratios for each pair of  
isotopes $(i,\,j)$ among (Xe, Ge, Te). The diagonal dotted line corresponds to  $M_{N,i}/M_{\nu,i}=M_{N,j}/M_{\nu,j}$.
Each point refers to one of the fifteen NME sets in Table~\ref{Tab:NME}.
Different markers label different models (SM, QRPA, EDF, IBM). 
} \end{minipage}
\end{figure}

Figure~\ref{Fig_01} shows the heavy-to-light NME ratios for each pair $(i,\,j)$ of isotopes among  (Xe, Ge, Te). 
The scatter plots show a significant spread (about a factor of three in each coordinate), that reflects 
the still large theoretical uncertainties affecting nuclear model calculations. The points also
tend to cluster around the diagonal lines, where $M_{N,i}/M_{\nu,i}=M_{N,j}/M_{\nu,j}$ and the two $0\nu\beta\beta$ decay
mechanisms become degenerate. A nearly degenerate situation occurs for the IBM cases in all $(i,\,j)$ pairs: 
in such cases, even precise measurements of $0\nu\beta\beta$ decay signals $(S_i,\, S_j)$ would not be able to
separate the contributions of light or heavy neutrino exchange via Eq.~(\ref{System}). Conversely, some 
QRPA and SM cases happen to be significantly off-diagonal for at least one  $(i,\,j)$ pair. In such cases, provided
that the (unquantified) model uncertainties are small enough not to cross the diagonal line,
the relative weight of the two mechanisms could be determined---at least in principle---via high-statistics  $(S_i,\, S_j)$ data.
The EDF case provides an intermediate situation, nearly degenerate for the (Xe,~Te) pair, and slightly nondegenerate for
the other two isotopic pairs. Since we do not know which model is close to the ``true'' NME values, we must currently accept 
the occurrence of all possibilities about the (non)degeneracy of the light and heavy neutrino mechanisms in 
$0\nu\beta\beta$ searches using (Xe, Ge, Te) data. However, it is interesting to note that while the (Xe,~Te) and (Xe,~Ge) points are scattered
on both sides of the diagonal line and along it, the (Ge,~Te) ones lie only on the upper side.
If this fact were not accidental, but suggestive of a model-independent inequality of the kind
$(M_N/M_\nu)_{\rm{Te}}>(M_N/M_\nu)_{\rm{Ge}}$, then the relative amount of decays mediated 
by light and heavy neutrinos could be determined in principle, the better the stronger the deviation from the
diagonal line.

\begin{figure}[t!]
\begin{minipage}[c]{0.91\textwidth}
\includegraphics[width=0.91\textwidth]{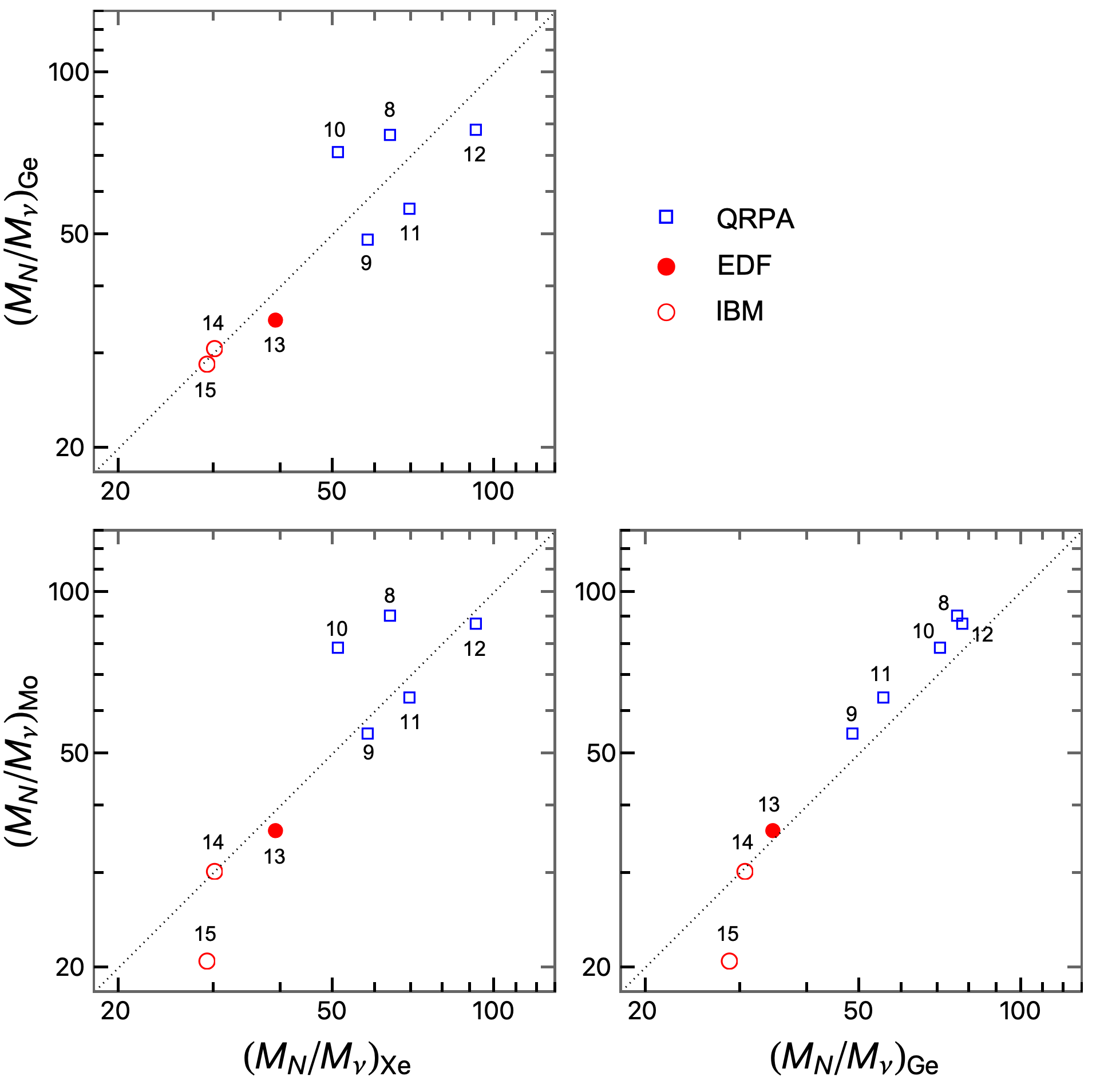}
\vspace*{-2mm}
\caption{\label{Fig_02}
\footnotesize As in Fig.~\ref{Fig_01}, but for each pair of isotopes among (Xe, Ge, Mo), and for the NME sets numbered in Table~\ref{Tab:NME}
from  8 to 15, as reported near each point. 
} \end{minipage}
\end{figure}

Figure~\ref{Fig_02} shows the heavy-to-light NME ratios for each pair $(i,\,j)$  among the (Xe, Ge, Mo) isotopes,
relevant for future ton-scale projects in nEXO, LEGEND, and CUPID.  One can make considerations
similar to Fig.~\ref{Fig_01} about the overall scatter of points (large, two-sided or one-sided) and 
about the occurrence of (non)degenerate cases. For later purposes, each point is distinguished
by the same NME set number reported in Table~\ref{Tab:NME} (first column).

A few comments are in order, about the issues of $g_A$ quenching \cite{Suhonen:2017krv} and additional short-range contributions
\cite{Cirigliano:2018hja,Jokiniemi:2021qqv} in $0\nu\beta\beta$ decay.
It is still matter of debate if, in $0\nu\beta\beta$ decay, 
the bare value $g_A=1.276$ should be effectively quenched by an isotope-dependent factor $q<1$ as in other observed weak-interaction processes, 
reducing NMEs dominated by axial-vector components by $\sim q^2$, 
and thus the expected decay rates by $\sim q^4$. 
It may be expected that quenching effects, if any, largely cancel in the NME ratios $(M_{N,i}/M_{\nu,i})$ that govern 
the (non)degeneracy of light and heavy neutrino mechanisms. This can be explicitly verified for some NME calculations in Table~\ref{Tab:NME} 
(numbered as 8--12, 14, and 15), where the axial NME components have been separately reported. For a typical 
quenching factor $q=0.79$ (leading to $qg_A\simeq 1)$ \cite{Suhonen:2017krv}, we find that the NME ratios
are generally altered by $\lesssim 10\%$ (and only in one case by 15\%), namely, by much less than the overall spread in Figs.~\ref{Fig_01}
and \ref{Fig_02}.   
Therefore, in the presence of quenching, the qualitative results of our degeneracy 
analysis would not be significantly altered, although the quantitative bounds on the effective parameters $m_\nu$ and $m_N$ would be weakened.
In the absence of clear indications about quenching effects in $0\nu\beta\beta$ decay, we assume $q=1$ for the sake of simplicity;
see also related comments in \cite{Lisi:2022nka}. 
\textcolor{black}{Concerning recently discussed short-range effects (whose size and sign
are still uncertain \cite{Pompa:2023jxc}), for different mechanisms they may contribute in different ways, not
cancelling in NME ratios \cite{Agostini:2022bjh}. While waiting for future progress on these issues, for the purposes of this work  
we surmise that the current factor-of-three spread of NMEs and of their ratios, as reported above,
is already large enough to qualitatively cover the effects of ``uncertain uncertainties,'' such as those  
related to $g_A$ quenching and short-range contributions.}

We conclude this section by completing the notation related to Eq.~(\ref{Tlightheavy}). As in
\cite{Lisi:2022nka}, we use the following units:
\begin{eqnarray}
\left[m_\nu\right] &=& \mathrm{meV}\ ,\label{mnuunit}\\ 
\left[m_N\right] &=& \mathrm{meV}\ ,\label{mNunit}\\ 
\left[T_i\right] &=& 10^{26}\, y \ ,\label{Tunit}\\
\left[S_i\right] &=& 10^{-26}\, y^{-1}\ ,\label{Sunit}\\
\left[G_i\right] &=& 10^{-26}\, y^{-1}\, (\mathrm{meV})^{-2}\ .\label{Gunit}
\end{eqnarray}
The phase space factors $G_i$ are taken from \cite{Deppisch:2020ztt} and, in our notation and units, they read:
\begin{eqnarray}
G_\mathrm{Xe} &=& 14.78 \times 10^{-6}\ , \\
G_\mathrm{Ge} &=& \phantom{0}2.40 \times 10^{-6}\ ,\\
G_\mathrm{Te} &=& 14.42 \times 10^{-6}\ ,\\
G_\mathrm{Mo} &=& 16.15 \times 10^{-6}\ .
\end{eqnarray}
The $G_i$ uncertainties are negligible in the context of this study.

\section{Separate and combined bounds from current data}
\label{Sec:Current}

In this Section we analyze the constraints on light and heavy effective neutrino masses, $m_\nu$ and $m_N$, 
as obtained by using the latest data from 
KamLAND-Zen \cite{KamLAND-Zen:2022tow} and EXO \cite{Anton:2019wmi} (Xe), GERDA \cite{Agostini:2020xta}
and MAJORANA \cite{Majorana:2022udl} (Ge), and CUORE \cite{CUORE:2021mvw} (Te).

\begin{table}[t!]
\centering
\resizebox{.9\textwidth}{!}{\begin{minipage}{\textwidth}
\caption{\label{tab:abc} 
Coefficients of the  $\Delta\chi^2_i$ function in Eq.~(\ref{Delta}), listed 
according to the isotopes in the first column and the (combinations of) experiments in the second column. 
The next three columns report our evaluation of the 
coefficients $(a_i,\,b_i,\,c_i)$ for separate experiments (upper five rows) and for their combinations
in the same isotope (lower three rows).  
 The sixth column reports our estimated 90\% C.L.\ half-life limits $T_{90}$, to be compared with the official one in the seventh column
(as taken from the reference in the eighth column, when applicabile).  
}
\begin{ruledtabular}
\begin{tabular}{rlrrrccc}
Isotope 	& Experiment or combination 		& $a_i~~~$ 	& $b_i~~~$ 	& $c_i~~~$ 	& $T_{90}/10^{26}\,\mathrm{y}$ & $T_{90}$ (expt.)
										& Reference \\ 
\hline
$^{136}$Xe	& KamLAND-Zen 				& 5.157		& 3.978		& 0.000		& 2.300 & 2.3
										& \cite{KamLAND-Zen:2022tow} \\
$^{136}$Xe	& EXO						& 0.440		& $-0.338$ 	& 0.065		& 0.350 & 0.35 
										& \cite{Anton:2019wmi}\\
$^{76}$Ge	& GERDA						& 0.000 	& \textcolor{black}{4.867} 	& 0.000 	& 1.800	& 1.8
										& \cite{Agostini:2020xta}\\
$^{76}$Ge	& MAJORANA					& 0.000		& \textcolor{black}{2.246} 	& 0.000		& 0.830 & 0.83 
										& \cite{Majorana:2022udl}\\
$^{130}$Te	& CUORE						& 0.245		& $-0.637$ 	& 0.414		& 0.216 & 0.22 
										& \cite{CUORE:2021mvw}\\
\hline
$^{136}$Xe	& Xe  (KamLAND-Zen + EXO)			
										& 5.597		& 3.640		& 0.000		& 2.260 & \textemdash 
										& \textemdash \\
$^{76}$Ge	& Ge (GERDA + MAJORANA)		& 0.000		& 7.113		& 0.000		& 2.629 & \textemdash 
										& \textemdash  \\
$^{130}$Te	& Te (CUORE, as above)& 0.245		& $-0.637$ 	& 0.414		& 0.216 &  0.22 
										& \cite{CUORE:2021mvw} \\
\end{tabular}
\end{ruledtabular}
\vspace*{-2mm}
\end{minipage}}
\end{table}

\vspace*{-1mm}
\subsection{Statistical data analysis}
\vspace*{-2mm}

We follow the  approach discussed in \cite{Capozzi:2021fjo,Lisi:2022nka}, by associating to each experiment a $\Delta\chi^2_i$
function of the form 
\begin{equation}
\Delta \chi^2_i(S_i) = a_i S^2_i + b_i S_i + c_i\ ,
\label{Delta}
\end{equation}
where the signal strength $S$ is the inverse of the half-life $T$, see Eq.~(\ref{Tlightheavy}). Combination of data are obtained by summing up the $\Delta\chi^2$'s.
The usual 90\% C.L.\ lower limit on $T$ ($T_{90}$) is obtained by setting $\Delta\chi^2=2.706$.  

Table~\ref{tab:abc},
updated from \cite{Lisi:2022nka} with the inclusion of the final MAJORANA results \cite{Majorana:2022udl},
reports the numerical values of the $(a_i,\,b_i,\,c_i)$ coefficients for the quoted experiments, as well as for their combination
in the same isotope.%
\footnote{Updated MAJORANA constraints are also quoted in \protect\cite{Pompa:2023jxc} within the same parametrization. 
For KamLAND-Zen, our coefficients are based on a fit to the official $\Delta\chi^2$ profile published 
in the supplemental material of \protect\cite{KamLAND-Zen:2022tow}, and are unchanged from those
reported in \protect\cite{Lisi:2022nka}. We note that the KamLAND-Zen coefficients in \protect\cite{Pompa:2023jxc}
are different from ours, although the same value of $T_{90}$ is recovered.}  
 It can be noticed the combined GERDA~+~MAJORANA limit on the Ge half-life,
 $T_{90}({\rm Ge})=2.629 \times 10^{26}$~y, is higher than the KamLAND-Zen~+~EXO limit on the 
Xe half-life, $T_{90}({\rm Xe})=2.260 \times 10^{26}$~y.

\begin{figure}[b!]
\begin{minipage}[c]{0.8\textwidth}
\includegraphics[width=0.85\textwidth]{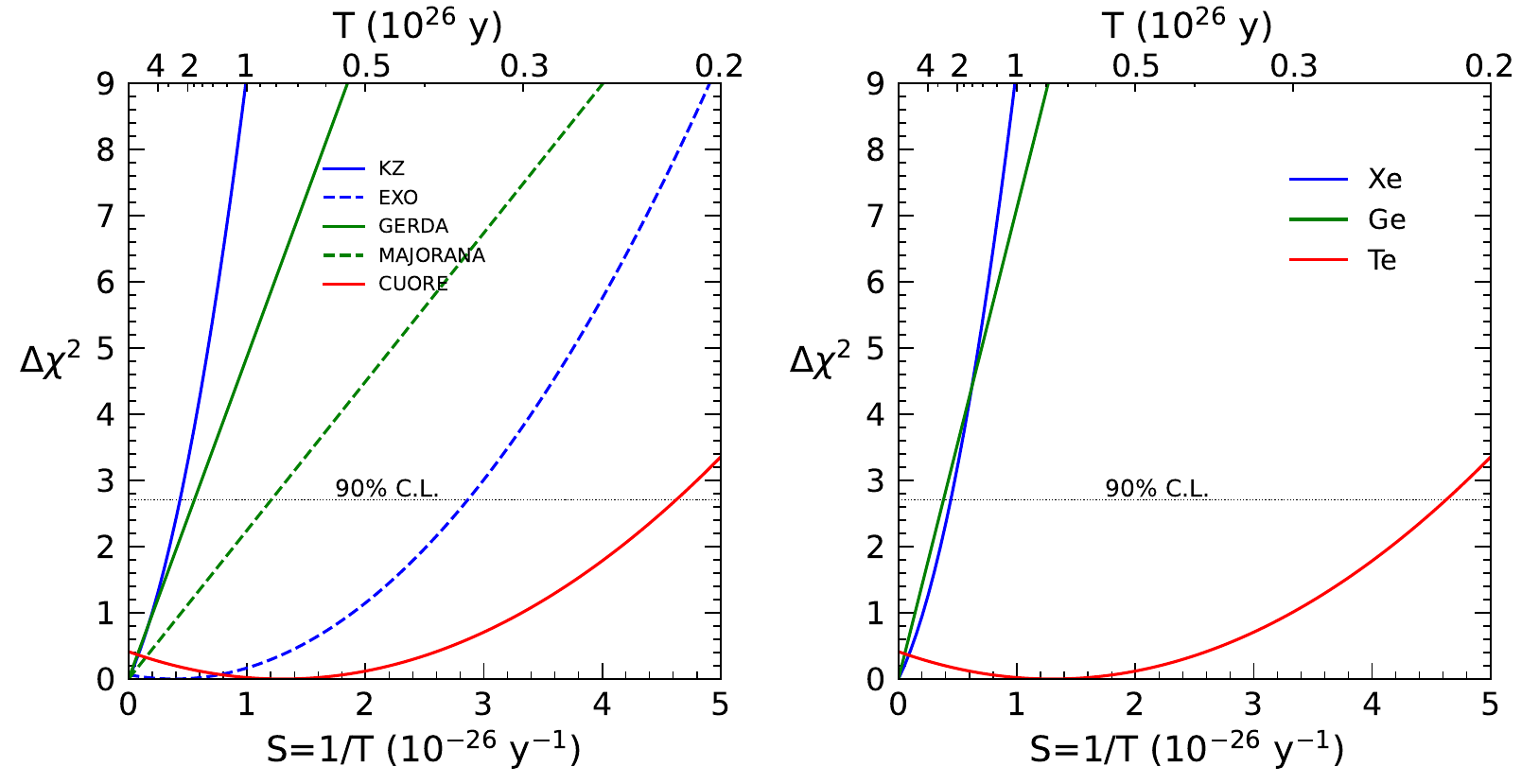}
\vspace*{-2mm}
\caption{\label{Fig_03}
\footnotesize $\Delta\chi^2$ functions in terms of the half-life $T$ (top abscissa) and of 
the signal strength $S = 1/T$ (bottom abscissa). Left and right panels: separate experiments and their combinations for the same isotope, respectively. Dotted horizontal lines intersect the curves at 90\% C.L. See the text for details.
} \end{minipage}
\end{figure}

Figure~\ref{Fig_03} shows the numerical information of Table~\ref{tab:abc} in graphical form; the left and right panels 
refer to separate experiments and to same-isotope combinations, respectively. 
Focussing on the right panel, it should be noted that: ($i$) For Ge and Xe, it is $\Delta\chi^2=0$ at null signal, while for Te there is a weak preference for a nonzero signal; ($ii$) as mentioned, the Xe constraints on $T$ are slightly weaker than those from Ge at 90\% C.L.; however, they become 
comparatively stronger for $\Delta\chi^2> 4.4$; ($iii$) in particular, at $3\sigma$, $T_{90}({\rm Ge}) \simeq 0.8 \times 10^{26}$~y, while
$T_{90}({\rm Xe}) \simeq 1 \times 10^{26}$. As emphasized in \cite{Capozzi:2021fjo,Lisi:2022nka}, there
is a lot more information in the $\Delta\chi^2$ functions than can be captured by the parameters $T_{90}$, often used to
characterize experimental performances.

In the following, bounds on the $0\nu\beta\beta$ effective parameters $m_\nu$ and $m_N$, both separately and in combination, will
be obtained by summing up the  $\Delta\chi^2_i$ associated to the 
Xe, Ge and Te signals $S_i$ appearing in Eq.~(\ref{Tlightheavy}). 
For the sake of simplicity, we shall present numerical and graphical 
bounds only for a reference C.L.\ of $2\sigma$
($\Delta\chi^2=4$).

\subsection{Light neutrino exchange only: $m_\nu\geq 0$, $m_N=0$}

The case of  light Majorana neutrino exchange only has a specific interest, being the simplest and most natural scenario 
for $0\nu\beta\beta$ decay. We update the recent analysis in \cite{Lisi:2022nka}, to account  for the latest Ge data and
for some differences in the adopted NME sets.%
\footnote{With respect to 
\protect\cite{Lisi:2022nka}, we have added some NME calculations 
from studies covering both light and heavy neutrino exchange, while we have excluded
those not covering the latter case. The overall number of adopted NME sets (fifteen) is accidentally the same as in \protect\cite{Lisi:2022nka}.}

Table~\ref{tab:Lightonly} reports, in the upper half, the $2\sigma$ upper bounds on the effective light Majorana mass $m_\nu$, 
for each of the fifteen representative NME sets $M_{\nu,i}$ 
($i=$~Xe, Ge, Te) listed in Table~\ref{Tab:NME}. Best-fit values of $m_\nu$ are reported in the lower half.
Concerning constraints from single isotopes,
in most cases Xe sets the strongest $2\sigma$ bounds, followed by weaker ones from Ge and Te. 
However, for the cases numbered as 6 and 7 (QRPA), the NME for Xe are the lowest, and the Ge 
bounds prevail on those from Xe and Te (comparable). Concerning the best fits for
single isotopes, only Te data favors $m_\nu>0$, 
due to the slight preference for a nonzero signal in Fig.~\ref{Fig_03}, in contrast with Xe and Ge data. The combination
of any two isotopes generally provides a bound stronger that the separate ones, except for the
cases involving Te with relatively large NME; in such cases, the joint bounds of Te with Ge or Xe are weakened, as a result
of a slight tension between the two isotopic data in terms of preferred $m_\nu$. This effect is more evident for noted NME sets 6 and 7, 
where the preference for $m_\nu>0$ at best fit persists in the Xe+Te combination. The global Xe+Ge+Te combination provides rather stable results for $m_\nu$, 
 characterized by $m_\nu=0$ at best-fit values and by upper bounds $m_{\nu,2\sigma}$ stronger than any
separate bound, for any choice of the NME set in Table~\ref{tab:Lightonly}. The spread of NME values 
implies a relatively large range for the corresponding $2\sigma$ bounds, 
\begin{equation}
\textcolor{black}{m_{\nu}\leq m_{\nu,2\sigma} \in   [43.1,\, 127.9] ~\mathrm{meV\ (Xe+Ge+Te) }\ ,}
\label{summarylight} 
\end{equation}
that unfortunately spans a factor of three. A significant reduction of these uncertainties is expected in the future, 
as a results of worldwide efforts in improving the NME calculations and benchmarking the nuclear models \cite{Cirigliano:2022oqy,Cirigliano:2022rmf}.

\begin{table}[t!]
\centering
\resizebox{.96\textwidth}{!}
{\begin{minipage}{1.0\textwidth}
\caption{\label{tab:Lightonly} \footnotesize 
Case with only light Majorana neutrino exchange ($m_\nu\geq 0$, $m_N=0$). Upper half: Bounds on $m_{\nu}$ in meV at $2\sigma$ level from current Xe, Ge, and Te data, both
separately and in combination, for each of the 15 representative NME sets listed in Tab.~\ref{Tab:NME}. 
Lower half:  Corresponding best-fit values of $m_{\nu} $, in meV. 
}
\begin{ruledtabular}
\begin{tabular}{lrrrrrrrrrrrrrrr}

Data~$\downarrow$ \textbackslash\  NME~$\rightarrow$ 			
            &   1  	&  2   &     3   &   4  &   5  &      6   &   7   &   8   &  9  &    10 &   11 &    12  &  13  &   14 &   15 \\  	
\hline
Xe			&  86.9 &  80.9  & 121.6 & 112.6 &  90.5 & 178.6 &  168.0 &  68.1 &  72.1 &  59.0 &   90.9 &  80.6 &  46.7 &  61.0 &   58.3 \\
Ge			& 167.5	& 157.7  & 143.6 & 135.6 & 172.3 & 155.1 &  142.4 &  92.0 &  89.0 &  83.2 &   93.8 &  87.1 &  80.1 &  94.2 &   76.4 \\
Te			& 220.5	& 205.6  & 340.0 & 315.3 & 229.7 & 209.9 &  189.0 & 152.1 & 145.6 & 129.5 &  156.5 & 139.3 & 124.5 & 153.7 &  146.6 \\
Xe+Ge		&  79.6	&  74.2  &  97.7 &  91.1 &  82.7 & 123.8 &  114.8 &  57.4 &  58.9 &  50.3 &   69.0 &  62.4 &  41.9 &  51.3 &   48.6 \\
Xe+Te		&  89.6	&  83.4  & 124.9 & 115.7 &  93.3 & 167.5 &  154.5 &  70.4 &  74.5 &  61.0 &   93.4 &  82.8 &  48.1 &  62.9 &   60.1 \\
Ge+Te		& 167.8	& 152.7  & 150.6 & 142.3 & 168.2 & 152.8 &  138.8 &  95.0 &  91.7 &  84.9 &   97.0 &  89.4 &  81.8 &  97.0 &   80.0 \\
Xe+Ge+Te	&  82.0	&  76.5  & 100.1 &  93.4 &  85.2 & 127.9 &  118.2 &  59.4 &  61.1 &  52.1 &   71.9 &  65.0 &  43.1 &  55.0 &   50.1 \\
\hline
Xe			&     0&     0&     0&     0&     0&     0&     0&     0&     0&     0&     0&     0&     0&     0&     0\\
Ge			&     0&     0&     0&     0&     0&     0&     0&     0&     0&     0&     0&     0&     0&     0&     0\\
Te			& 108.8& 101.4& 167.7& 155.6&  113.3& 103.5&  93.3&  75.1&  71.8&  63.9&  77.2&  68.7&  61.4&  75.8&    72.4\\
Xe+Ge		&     0&     0&     0&     0&     0&     0&     0&     0&     0&     0&     0&     0&     0&     0&     0\\
Xe+Te		&     0&     0&     0&     0&     0&  31.7&  36.0&     0&     0&     0&     0&     0&     0&     0&     0\\
Ge+Te		&     0&     0&     0&     0&     0&     0&     0&     0&     0&     0&     0&     0&     0&     0&     0\\
Xe+Ge+Te	&     0&     0&     0&     0&     0&     0&     0&     0&     0&     0&     0&     0&     0&     0&     0\\
\end{tabular}
\end{ruledtabular}
\end{minipage}}
\end{table}

\subsection{Heavy neutrino exchange only: $m_\nu=0$, $m_N\geq 0$}

The alternative case of $0\nu\beta\beta$ decay mediated only by heavy Majorana neutrinos (or by any other nonstandard mechanism) 
cannot be excluded a priori, e.g.\ if a decay 
signal is observed, but the three complex terms in Eq.~(\ref{mlight}) interfere destructively and lead to $m_\nu\simeq 0$
(an allowed scenario for normal ordering of neutrino masses).

Table~\ref{tab:Heavyonly} reports the $2\sigma$ upper limits and best fits for the effective heavy Majorana mass $m_N$, in the same format used in  Table~\ref{tab:Lightonly}
for $m_\nu$. According to Eq.~\ref{Tlightheavy}, the $m_N$ bounds are expected to differ from the $m_\nu$ bounds by typical NME ratios, namely, by a factor 
$M_N/M_\nu\simeq 30$---$90$  (see Fig.~\ref{Fig_01}), as it indeed occurs numerically. Apart from this overall scaling of the bounds, the previous comments
about the impact of separate and combined isotopic data in Table~\ref{tab:Lightonly} also apply to Table~\ref{tab:Heavyonly}. The results on the $2\sigma$ 
upper bounds can be
summarized as 
\begin{equation}
\textcolor{black}{m_{N}\leq m_{N,2\sigma} \in   [0.75,\, 2.1] ~\mathrm{meV\ (Xe+Ge+Te) }\ .}
\label{summaryheavy} 
\end{equation}

\begin{table}[t!]
\centering
\resizebox{.96\textwidth}{!}
{\begin{minipage}{1.0\textwidth}
\caption{\label{tab:Heavyonly} \footnotesize 
Case with only heavy Majorana neutrino exchange ($m_\nu= 0$, $m_N\geq 0$). Upper half: Bounds on $m_{N}$ in meV at $2\sigma$ level from current Xe, Ge, and Te data, both
separately and in combination, for each of the 15 representative NME sets listed in Tab.~\ref{Tab:NME}. 
Lower half:  Corresponding best-fit values of $m_{N} $, in meV. 
}
\begin{ruledtabular}
\begin{tabular}{lrrrrrrrrrrrrrrr}

Data~$\downarrow$ \textbackslash\  NME~$\rightarrow$
  			&   1  &     2   &      3 &     4  &      5  &      6  &     7   &   8   &    9  &     10 &      11 &      12  &    13 &      14 &       15 \\  	
\hline
Xe			& 1.71  &  1.19 &  2.01 &   1.39 &    1.73 &    2.96 &   2.19 &   1.06 &   1.24 &   1.15 &    1.30 &   0.87 &   1.19 &  2.02 &   2.00 \\
Ge		    & 3.72  &  2.58 &  3.84 &   2.40 &    3.65 &    2.58 &   1.65 &   1.21 &   1.83 &   1.75 &    1.69 &   1.12 &   2.32 &  3.08 &   2.67 \\
Te			& 4.17  &  2.90 &  6.44 &   4.48 &    4.22 &    3.18 &   2.01 &   1.80 &   2.54 &   1.58 &    2.31 &   1.52 &   3.14 &  4.87 &   4.88 \\
Xe+Ge		& 1.59  &  1.11 &  1.84 &   1.24 &    1.60 &    2.06 &   1.39 &   0.84 &   1.07 &   0.87 &    1.08 &   0.72 &   1.09 &  1.76 &   1.68 \\
Xe+Te		& 1.76  &  1.22 &  2.05 &   1.42 &    1.78 &    2.66 &   1.79 &   1.09 &   1.28 &   1.14 &    1.34 &   0.89 &   1.23 &  2.09 &   2.06 \\
Ge+Te		& 3.36  &  2.33 &  3.98 &   2.51 &    3.35 &    2.44 &   1.55 &   1.22 &   1.81 &   1.15 &    1.66 &   1.10 &   2.28 &  3.15 &   2.78    \\
Xe+Ge+Te	& 1.65  &  1.14 &  1.88 &   1.27 &    1.65 &    2.10 &   1.39 &   0.88 &   1.11 &   0.90 &    1.13 &   0.75 &   1.13 &  1.82 &   1.73 \\
\hline
Xe			&     0&     0&     0&     0&     0&     0&     0&     0&     0&     0&     0&     0&     0&     0&     0\\
Ge			&     0&     0&     0&     0&     0&     0&     0&     0&     0&     0&     0&     0&     0&     0&     0\\
Te			& 2.06 &   1.43&   3.18&   2.21&   2.08&   1.57&   0.99&   0.89&   1.25&   0.78&   1.14&   0.75&   1.55&   2.40&   2.37\\
Xe+Ge		&     0&     0&     0&     0&     0&     0&     0&     0&     0&     0&     0&     0&     0&     0&     0\\
Xe+Te		&     0&     0&     0&     0&     0&   0.99&   0.87&     0&     0&     0&     0&     0&     0&     0&     0\\
Ge+Te		&     0&     0&     0&     0&     0&     0&     0&     0&     0&     0&     0&     0&     0&     0&     0\\
Xe+Ge+Te	&     0&     0&     0&     0&     0&     0&     0&     0&     0&     0&     0&     0&     0&     0&     0\\
\end{tabular}
\end{ruledtabular}
\end{minipage}}
\end{table}

Effective heavy Majorana masses at a scale comparable to these bounds, $m_N \sim O(1)$~meV, can be realized via Eq.~\ref{mheavy} in LR-symmetric theories
assuming favorable physics scales, such as $m_{W_R}\sim $~few~TeV, $M_h\in O(10^{2\pm 1})$~GeV and $V_{eh}\sim O(U_{ei})$;
see \cite{Patra:2023ltl} for a recent model construction, that will be discussed later in more detail. In general, such models
may allow comparable $0\nu\beta\beta$ contributions from both $m_\nu$ and $m_N$, which is the next case to be considered.

\vspace*{1mm}
\subsection{Non-interfering light and heavy neutrinos: $m_\nu\geq 0$, $m_N\geq 0$}
\vspace*{1mm}

\begin{figure}[t!]
\begin{minipage}[c]{0.91\textwidth}
\includegraphics[width=0.91\textwidth]{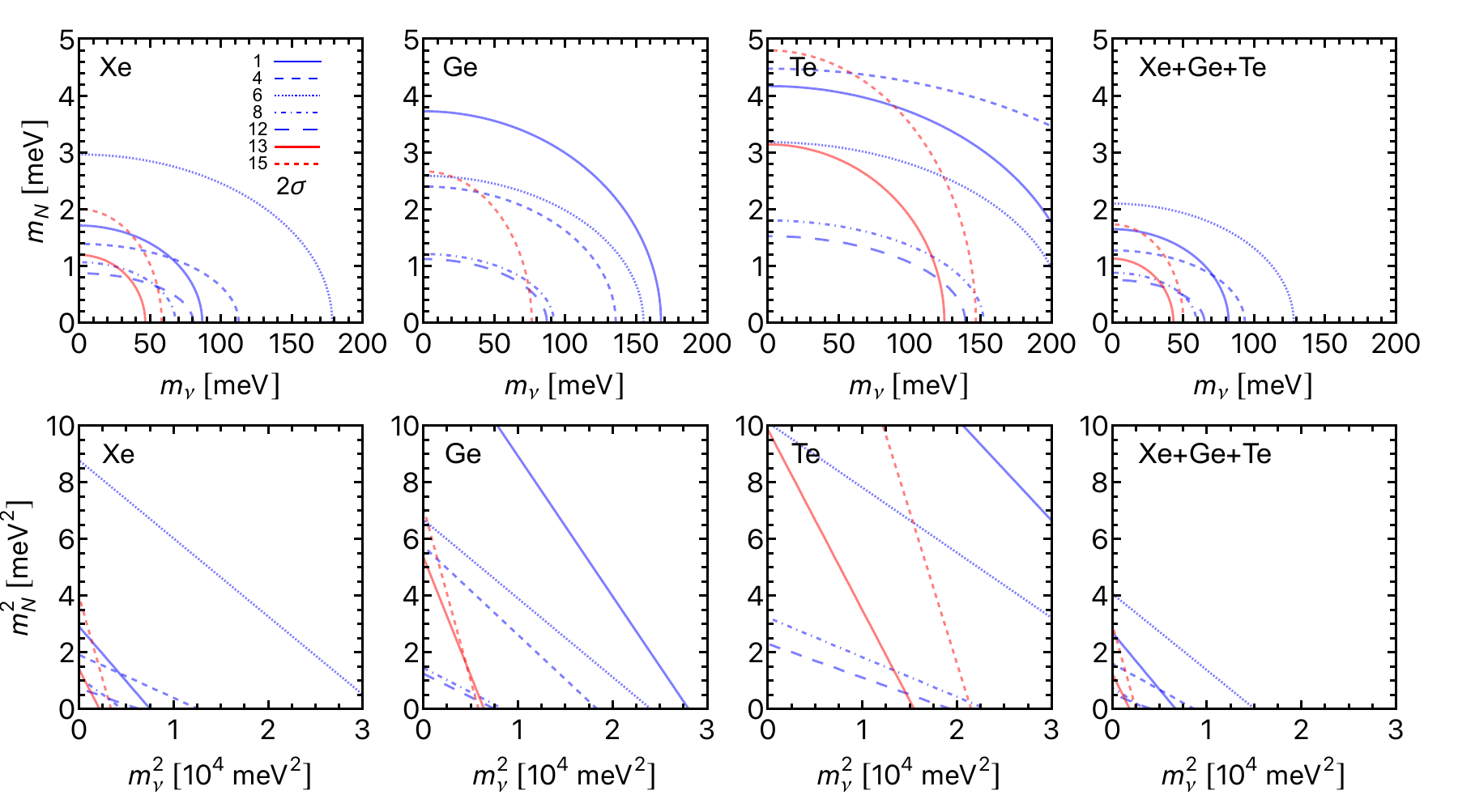}
\vspace*{-2mm}
\caption{\label{Fig_04}
\footnotesize Joint upper bounds on the effective Majorana masses for the exchange of light neutrinos ($m_\nu$) and heavy neutrinos ($m_N$) from current Xe, Ge and Te data, 
under the assumption of non-interfering exchange, Eq.~(\ref{Tlightheavy}). The legenda with colored line types 
refers to representative NME sets, numbered  as in Table~\ref{Tab:NME}.
All bounds are derived at the $2\sigma$ confidence level 
($\Delta\chi^2=4.0$), and refer to the pairs $(m_\nu,\,m_N)$ and  $(m^2_\nu,\,m^2_N)$ in the upper and lower panels, respectively. 
From left to right, the panels refer first to the three separate Xe, Ge and Te bounds, and then to their combination Xe+Ge+Te. 
} \end{minipage}
\end{figure}

Figure~\ref{Fig_04} shows, in the upper panels, the joint upper bounds in terms of  the effective Majorana mass parameters $(m_\nu,\,m_N)$,
using separate and combined Xe, Ge and Te data from current experiments.  Regions below each curve are allowed at $2\sigma$  $(\Delta \chi^2=4)$. 
To avoid confusion, we show only selected cases with
relatively weak or strong bounds, for seven NME sets representative of the SM, QRPA, EDF and IBM models, numbered as in Table~\ref{Tab:NME}. 
In the limits $m_N=0$ and $m_\nu=0$, we recover the $2\sigma$ bounds reported in Tables~\ref{tab:Lightonly} and~\ref{tab:Heavyonly}, respectively.

The lower panels of Fig.~\ref{Fig_04} map the same bounds as in the upper panels, but in the squared variables $(m^2_\nu,\,m^2_N)$. Since 
Eq.~(\ref{Tlightheavy}) is linear in such variables, the bounds for separate Xe, Ge, Te isotopes
are exactly linear in such scales. For a given NME set,
the slope of the linear bound reflects the ratio $M_{ N,i}/M_{\nu,i}$ for the $i$-th isotope: the smaller the ratio, the steeper the slope. The bounds from
the Xe+Ge+Te combination stem from a best fit to a system of equations, and are not expected to be linear in principle (they should be arcs 
of ellipses in the squared variables). In practice, they 
turn out to be very close to linear, the combinations being typically dominated by a single isotope; see the rightmost lower panel in Fig.~\ref{Fig_04}.

We can thus summarize the joint $2\sigma$ bounds on $(m^2_\nu,\,m^2_N)$ in an approximately linear parametric form, applicable to any considered NME set, and 
smoothly interpolating between the 
squares of the $2\sigma$ limits $m_{\nu,2\sigma}$ and  $m_{N,2\sigma}$ reported for the Xe+Ge+Te combination in 
Tables~\ref{tab:Lightonly} and~\ref{tab:Heavyonly}, respectively: 
\begin{eqnarray}
m^2_\nu &\leq & (1-\alpha) m^2_{\nu,2\sigma} \label{param1}\ , \\
m^2_N &\leq & \alpha\, m^2_{N,2\sigma} \label{param2}  \ , 
\end{eqnarray}
where $\alpha\in [0,\,1]$. For $\alpha=0$ ($\alpha=1$), one recovers 
the separate bounds for the exchange of only light (heavy) Majorana neutrinos. 
The results discussed in this Section
represent the most updated bounds on non-interfering light and heavy Majorana neutrino exchange that can be derived from current
multi-isotope $0\nu\beta\beta$ data and for recent NME sets. 

Some comments are in order on  NME uncertainties. In this work, we choose to take the spread of the 
results, stemming from different NME sets, as a proxy for the
(largely unknown) theoretical uncertainties affecting nuclear models. In principle, more refined approaches are possible. For instance, within a nuclear model,
one could construct many variants, possibly constrained by pertinent data, 
and infer a probability distribution function (p.d.f.) for the NME,
accounting for covariances among isotopes. This approach was proposed for light Majorana neutrino exchange
in  \cite{Faessler:2008xj,Faessler:2013hz}, using QRPA model variants benchmarked by $2\nu\beta\beta$ data; see also the discussion in Sec.~III.B of \cite{Capozzi:2021fjo}.
Extensions to other exchange mechanisms within the same model involve further assumptions about the joint p.d.f.\ of the corresponding NME, see e.g.\ \cite{Ge:2017erv}.
Furthermore, to cover also different models one must assume that the global p.d.f.\ of the NME can be inferred from the spread of published values, see e.g.\ \cite{Agostini:2022bjh}.
At present, rather than relying on a chain of assumptions about the global NME (co)variances, we prefer to stick to the scattered central values 
of the NME in Table~\ref{Tab:NME}. Such a choice shall be reconsidered, if significant improvements are achieved on global NME p.d.f., covering multiple nuclear models, 
benchmark data,  isotopes, and decay mechanisms, as envisaged in \cite{Cirigliano:2022oqy,Cirigliano:2022rmf}.

Finally, when passing from real data (this Section) to prospective signals (next Section), it should be noted that the NME sets appear in two steps, namely,
in the generation of mock data (``true'' NME) and in their combined fit (``test'' NME). Given our ignorance of the NME set chosen by nature, we must
not only allow for different true NME's, but also for test NME's different from the true ones. This approach was followed, e.g., in 
\cite{Faessler:2011rv} to illustrate the degeneracy of multiple interfering mechanisms, and more recently in \cite{Pompa:2023jxc} to study the effect of short-range 
contributions to light Majorana neutrino exchange. We shall adopt a similar approach below, in selected prospective cases.

\vspace*{-2mm}
\section{Analysis of prospective signals in ton-scale experiments}
\label{Sec:Future}
\vspace*{-1mm}

In the previous Section, we studied upper bounds on $m_\nu$ and $m_N$ placed by current $0\nu\beta\beta$ data for a given NME set.
Should future data be consistent with no decay, the bounds would become stronger, but the qualitative aspects of our results would not change.
Of course, the analysis might become more interesting in the presence of observed decays. 

In this Section, we analyze selected examples of prospective $0\nu\beta\beta$ decay signals accessible at $>3\sigma$
in future ton-scale projects, with reference to nEXO (Xe), LEGEND (Ge) and CUPID (Mo), assuming 
a nominal exposure of 1 ton $\times$ 10 years, together with representative NME sets from Table~\ref{Tab:NME}. After setting the
statistical tools, we discuss the reconstruction
of  hypothetical signals for fixed NME sets, and then study the effect of swapping true and test NME sets.  
The bounds will be shown in the plane charted by the squared parameters $(m^2_\nu,\,m^2_N)$, in order to illustrate 
 the constraints from single isotopes and the (non)degeneracy effects among different isotopes, related to NME ratios.

\subsection{Statistical approach}

We follow the approach advocated in \cite{Agostini:2022zub}, where a generic $0\nu\beta\beta$ search is characterized, for a given exposure $\cal E$, as a simple counting experiment, observing a total number of $n$ events with respect to an average of $\mu$ events. For our purposes, the related (poissonian) $\chi^2$ function can be approximated as \cite{ParticleDataGroup:2022pth}:
\begin{equation}
\chi^2(n | \mu) \simeq 2 \left[ \mu - n + n \ln(n/\mu)\right]\ . \label{Poisson}
\end{equation}
In general, $n=n_B+n_S$, where $n_B$ counts the background events (assumed to be known) 
and $n_S$ the signal events. The so-called discovery sensitivity, i.e., 
the level of rejection of the background-only hypothesis when $n_S>0$, can be evaluated by taking 
$n=n_B+n_S$ and $\mu=n_B$. The so-called exclusion sensitivity, i.e., the level of rejection of a test signal $n_S$, if only $n_B$ events are observed, amounts to take 
$n=n_B$ and $\mu=n_B+n_S$. Finally, in the general case where a (true) signal $n_{\bar S}>0$ is assumed to be observed, the likelihood of an alternative (test) signal $n_S$ corresponds to take $n=n_B+n_{\bar S}$ and $\mu=n_B+n_S$ in the above $\chi^2$. 

Of course, in real $0\nu\beta\beta$ experiments, refined estimates for
the likelihood of background and signal events can include further measured or simulated
information (in terms of energy, time, and position), that is not contained in just two numbers ($n_S$ and $n_B$). 
However, by appropriate choices of effective values for $n_B$ at given exposures, one
can obtain a reasonable approximation for the sensitivity to a generic signal $n_S$ 
(see Table~IV in \cite{Agostini:2022zub} and Table~II in \cite{Agostini:2022bjh}), as also adopted by some experimental groups for prospective studies (see, e.g., Sec.\ V.D.9 in \cite{LEGEND:2021bnm}). 

Contact with our notation is obtained, for each isotope 
$i$, by 
expressing $n_{S_i}$ and $n_{B_i}$ in terms of the signal strength $S_i$ and associated background level $B_i$, respectively, via a common conversion factor $k_i$: 
\begin{eqnarray}
n_{S_i} &=& k_i\, {S_i}\ ,\\
n_{B_i} &=& k_i\, {B_i}\ .
\end{eqnarray}
For a given exposure ${\cal E}_i$, the value of $B_i$ (proportional to ${\cal E}_i)$ determines the $\chi^2$ function needed to test a signal $S_i$. In the general case mentioned above, i.e., 
assuming a priori a true signal $\bar S_i$, the $\chi^2_i$ for any test signal $S_i$ is given by 
\begin{equation}
\label{chi2future}
\chi^2_{i}(S_i) = {2}\,{k_i}\left[S_i-\bar S_i+(B_i+\bar S_i)\ln\left(\frac{B_i+\bar S_i}{B_i+S_i}\right)\right]\ .
\end{equation}
In particular, 
the assumption of a $3\sigma$ discovery signal $\bar S_i=S^{3\sigma}_i$ corresponds to have $\chi^2_i=9$ for $S_i=0$ (and, of course,
$\chi^2_i=0$ for $S_i=S^{3\sigma}_i$). A second-order expansion in $\delta = S_i-\bar S_i$ would provide
a quadratic form, as in Eq.~(\ref{Delta}).

In this work, the $B_i$ values have been obtained  by tuning the equivalent parameters proposed in
\cite{Agostini:2022zub,Agostini:2022bjh}, so as to optimize the agreement with the discovery sensitivity 
profiles presented for various exposures in the latest 1-ton design studies by
nEXO (Figs.~12 and 13 in \cite{nEXO:2021ujk}), LEGEND (Fig.~19 of \cite{LEGEND:2021bnm}), 
and CUPID (Fig.~2 in \cite{CUPID:2022wpt}), as well as with the corresponding 
$3\sigma$ discovery values for the half-life ($T^{3\sigma}_i=1/S^{3\sigma}_i$) quoted therein. 
Table~\ref{tab:Poisson} reports our reference values for the parameters $k_i$ and $B_i$, as well as for the associated ones 
$T^{3\sigma}_i$, $S^{3\sigma}_i$, $n_{B_i}=k_i B_i$ and $n^{3\sigma}_{S_i}=k_i S^{3\sigma}_i$, assuming for all 
experiments an isotopic mass of 1~ton and 10 years of data taking, corresponding to an exposure
\begin{equation}
\label{Exposure}
{\cal{E}}_i={\cal{E}}=10~\mathrm{ton \ y}\ .
\end{equation}

\begin{table}[t!]
\centering
\resizebox{.98\textwidth}{!}{\begin{minipage}{\textwidth}
\caption{\label{tab:Poisson} 
Poissonian $\chi^2$ for ton-scale experiments [Eq.~(\ref{chi2future})]: 
reference parameters $k_i$  and $B_i$  and associated quantities, for a common
exposure ${\cal E}=10$~ton~y; see the text for details.
For each isotope $i$, the $3\sigma$ discovery half-life $T^{3\sigma}_i$ matches the one 
in the available design studies (as quoted in the last column).}
\begin{ruledtabular}
\begin{tabular}{clccccccc}
Isotope   	& Project & $k_i$ & $B_i$ & $T_i^{3\sigma}$ & $S_i^{3\sigma}$ & $n_{B_i}$ & $n_{S_i}^{3\sigma}$		 & Ref. \\ 
  $i$            &         & [$10^{26}\, y$] & [$10^{-26}\,y^{-1}$] & [$10^{26}\,y$] & [$10^{-26}\,y^{-1}$] & \\ 
\hline
$^{136}$Xe	& nEXO 		& 622	& $8.84\times 10^{-3}$	& 74  & $13.5\times 10^{-3}$	& 5.50 & 8.41 & \cite{nEXO:2021ujk} \\
$^{76}$Ge	& LEGEND	& 403	& $0.99\times 10^{-3}$ 	& 130 & $7.69\times 10^{-3}$	& 0.40 & 3.10  & \cite{LEGEND:2021bnm}\\
$^{100}$Mo	& CUPID		& 315 	& $2.53\times 10^{-3}$ 	& 80  & $12.5\times 10^{-3}$ 	& 0.80 & 3.94 & \cite{CUPID:2022wpt}\\
\end{tabular}
\end{ruledtabular}
\end{minipage}}
\end{table}

\begin{figure}[h!]
\begin{minipage}[c]{0.8\textwidth}
\includegraphics[width=0.8\textwidth]{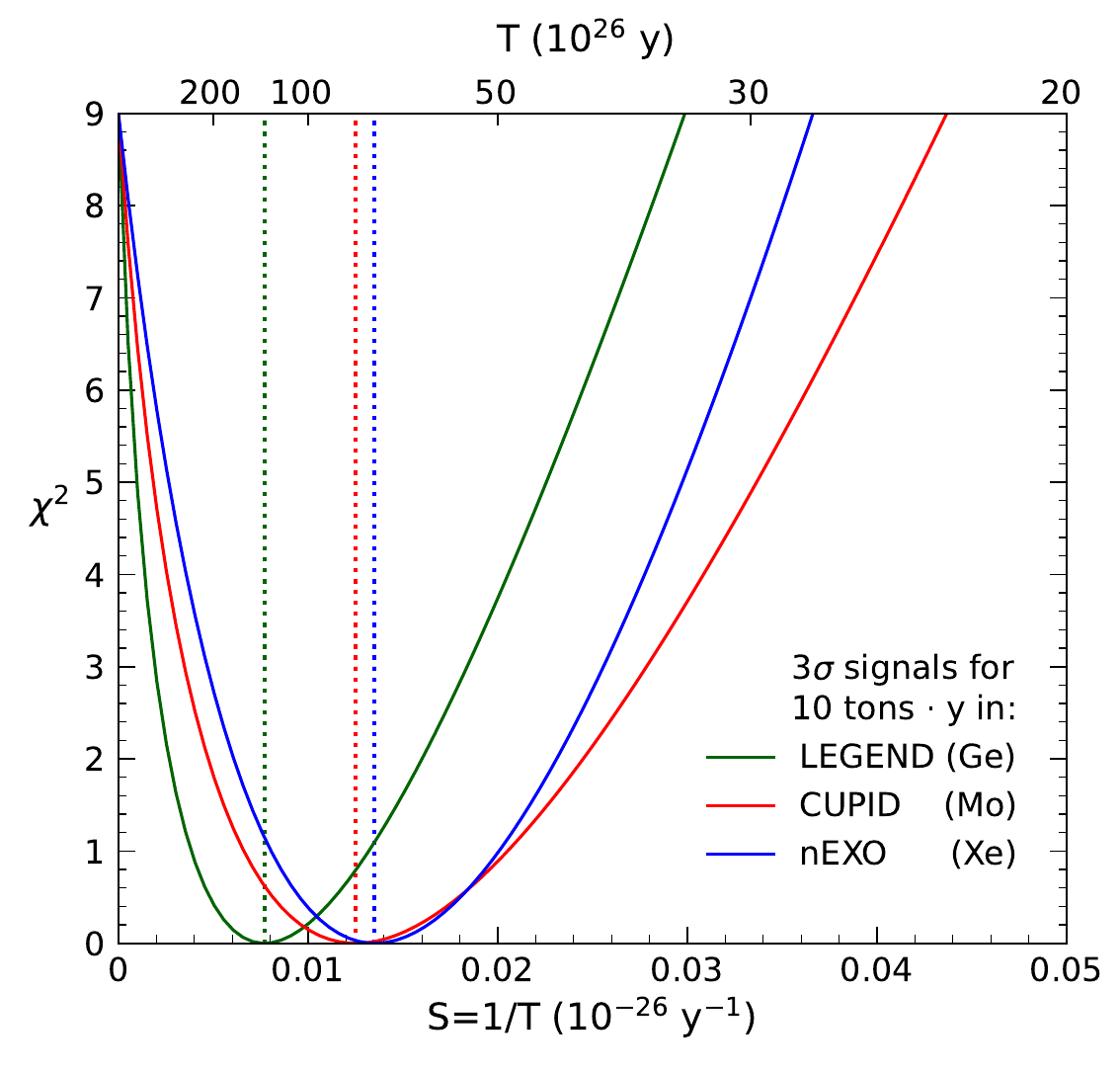}
\vspace*{-3mm}
\caption{\label{Fig_05}
\footnotesize Poissonian $\chi^2$ functions in terms of the half-life $T$ (top abscissa) and of 
the signal strength $S = 1/T$ (bottom abscissa), for an assumed $3\sigma$ discovery in
nEXO, LEGEND and CUPID, with an exposure of 10 ton~y. Each curve provides the best
fit ($\chi^2=0)$ at the corresponding discovery half-life $T^{3\sigma}$ (marked by a vertical dotted line)
and rejects the test case of null signal at $3\sigma$ ($\chi^2=9$).
} \end{minipage}
\end{figure}

Figure~\ref{Fig_05} shows the functions $\chi^2_i(S_i)$ for an assumed
$3\sigma$ prospective signal in nEXO, LEGEND, and CUPID,
as obtained from Eq.~(\ref{chi2future}) by setting $\bar S_i = S^{3\sigma}_i$. By construction, the 
best fit ($\chi^2_i=0$), marked by a vertical dotted line, 
is reached for $S_i= S_i^{3\sigma}$, while the null signal $S_i=0$  is rejected at
$3\sigma$   ($\chi^2_i=9$). With respect to Fig.~\ref{Fig_03} (current data), note the change 
of the scale in abscissa by two orders of magnitude, from typical values $T\sim O(10^{26})$~y to $\sim O(10^{28})$~y.

\vspace*{-3mm}
\subsection{Analysis with identical (true and test) NME sets}
\vspace*{-1mm}

For simplicity, we assume three representative pairs for the effective Majorana masses, 
\begin{equation}
\label{seed-m}
(\overline m_\nu,\,\overline m_N) = 
\left\{\begin{array}{ll}
(20,\,0)\ \mathrm{meV} & \leftarrow \mathrm{light}\ \nu_k\ \mathrm{only}\ ,\\
(0,\,0.4)\ \mathrm{meV} & \leftarrow\mathrm{heavy}\ N_h\ \mathrm{only}\ ,\\
(15,\,0.3)\  \mathrm{meV} & \leftarrow\mathrm{light}\ \nu_k + \mathrm{heavy}\ N_h\ .\\
\end{array}\right.
\end{equation}
The chosen values, for any NME set considered herein, are small enough to satisfy the most stringent $2\sigma$ upper bounds placed by current data,
and high enough to provide a $>3\sigma$ signal $\bar S_i$ in each ton-scale experiment,
where
\begin{equation}
\label{seed-S}
\bar S_i = G_i(M^2_{\nu,i}\, \overline m^2_\nu + M^2_{N,i}\, \overline m^2_N)\ .
\end{equation}
The true signals $\bar S_i$ are fitted by test signals $S_i$ via Eq.~(\ref{chi2future}), both separately and in combination. 
We start by making the futuristic assumption that the true and test NME are the same, as if they had no uncertainty; 
this assumption will be dropped in the next Section. As a consequence, the true values in Eq.~(\ref{seed-m})
are reconstructed as best-fit test values, with $\chi^2=\Sigma \chi^2_i=0$; what matters is the just the uncertainty of this reconstruction, that we show at the $2\sigma$ level ($\chi^2=4$). In the  ($m^2_\nu,m^2_N$) plane, 
$\chi^2$ isolines appear as slanted bands for separate isotopes, while they appear as ellipses
in multi-isotope combinations. The slopes of the bands are governed by the  $M_{N,i}/M_{\nu,i}$ ratios, so that their mutual
overlap (and thus the extension of the ellipse) depends of the differences among these ratios: the smaller the differences,
the closer the slopes, the larger the overlap, the higher the degeneracy between the two $0\nu\beta\beta$ mechanisms.

\begin{figure}[t]
\begin{minipage}[c]{0.99\textwidth}
\includegraphics[width=0.99\textwidth]{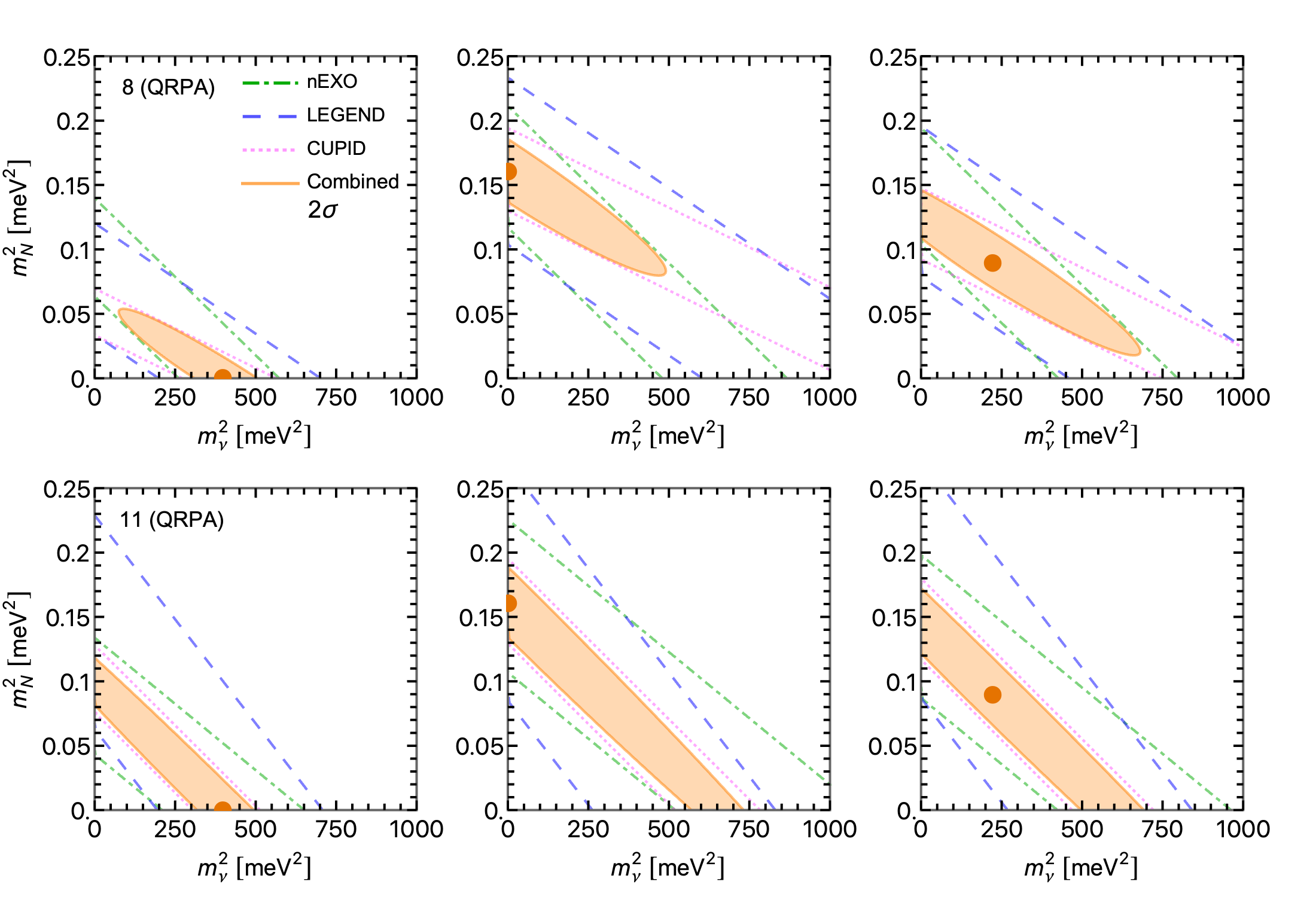}
\vspace*{-2mm}
\caption{\label{Fig_06}
\footnotesize Fit to prospective data from ton-scale projects (nEXO, LEGEND, CUPID), both separately (slanted bands) 
and in a global combination (ellipses), in the plane charted by the squared mass parameters $(m^2_\nu,\,m^2_N)$, at $2\sigma$ level.
The upper and lower panels refer to the NME sets numbered as 8 and 11 in Table~\ref{Tab:NME}. The left, middle 
and right panels refer to the three representative cases in Eq.~(\ref{seed-m}), identified by solid circles.
} \end{minipage}
\end{figure}

Among the pertinent NME sets numbered from 8 to 15 in Table~\ref{Tab:NME} and in Fig.~\ref{Fig_02}, we choose four representative
ones: the two QRPA sets labelled as 8 and 11, that provide relatively high ratios $M_{N_i}/M_{\nu,i}$, and appear on
opposite sides of the diagonal in two of the three planes of Fig.~\ref{Fig_02}; the  EDF set labelled as 13,
that provides intermediate values of the ratios $M_{N,i}/M_{\nu,i}$, and appears to be close to all diagonals 
in Fig.~\ref{Fig_02}; and the IBM set numbered as 15, that provides relatively low ratios $M_{N,i}/M_{\nu,i}$,
significantly off-diagonal in two of the three planes of Fig.~\ref{Fig_02}. The other NME sets would 
provide qualitatively similar results.

Figure~\ref{Fig_06} shows the $2\sigma$ constraints from ton-scale experiments 
for the three values of the mass parameters in Eq.~(\ref{seed-m})
(left, middle and right panels), using the QRPA NME set labelled as 8 (upper panels) and 11 (lower panels).
In the upper panels, the separate bands have quite different slopes, and their combination (an ellipse) 
allows to distinguish at least the extreme cases. In particular, for the true 
cases of only light (or heavy) neutrinos, the opposite test cases of only heavy (or light) neutrinos
are rejected at $>2\sigma$. For the case with both mechanisms at the same time (right panel),  
the limit $m_N=0$ is rejected, while $m_\nu=0$ is allowed, as a result of the relatively high ratio
$M_{N,i}/M_{\nu,i}$ in all isotopes. In the lower panels (NME set 11), the various
slopes are only moderately different, and the two mechanisms become effectively degenerate 
at the $2\sigma$ level: the allowed ellipse interpolates between the limiting cases and is
not able to separate them.

\begin{figure}[t!]
\begin{minipage}[c]{0.99\textwidth}
\includegraphics[width=0.99\textwidth]{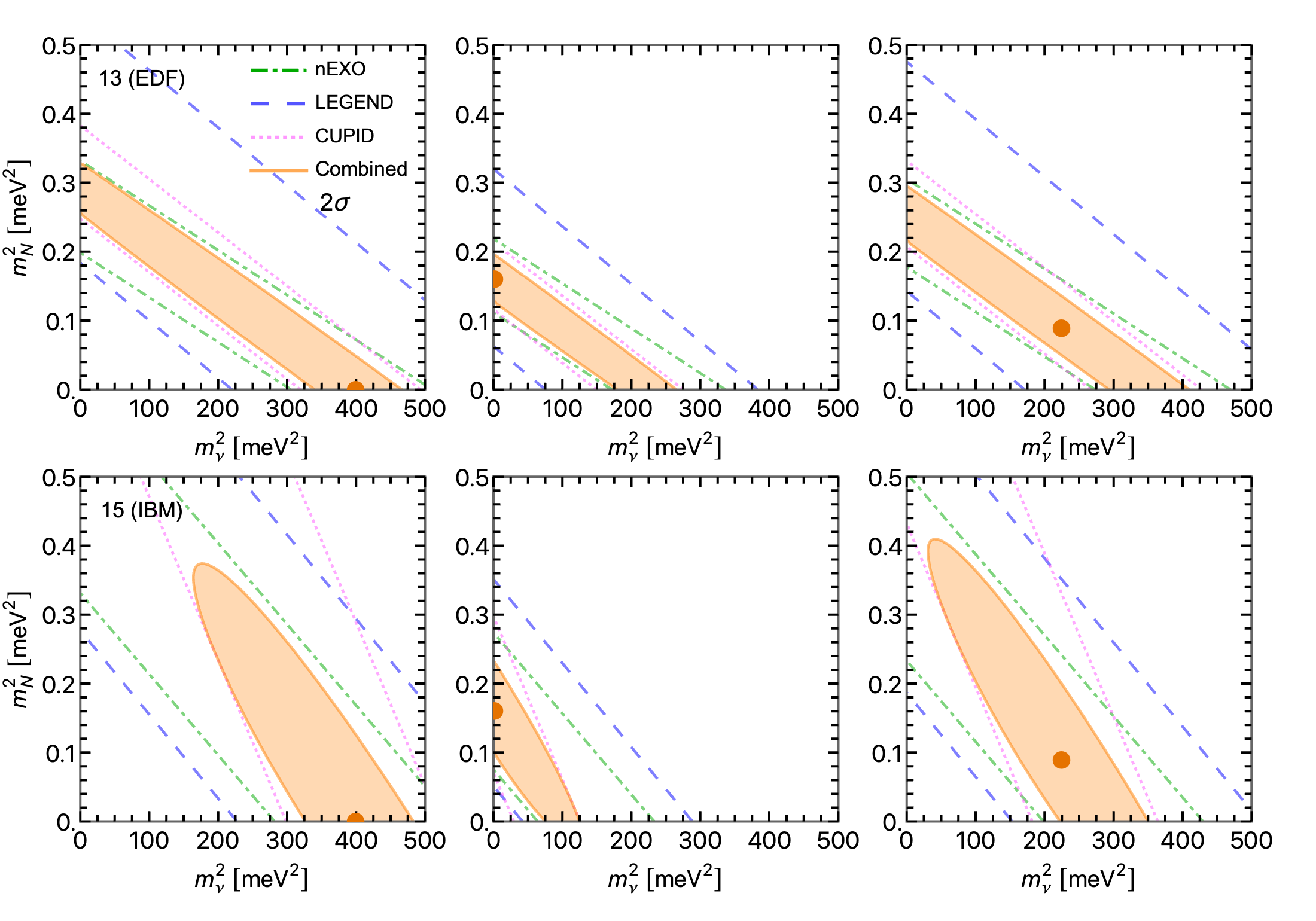}
\vspace*{-2mm}
\caption{\label{Fig_07}
\footnotesize As in Fig.~\ref{Fig_06}, but for the NME sets numbered as
as 13  and 15 in Table~\ref{Tab:NME}.
} \end{minipage}
\end{figure}

Figure~\ref{Fig_07} is analogous to Fig.~\ref{Fig_06}, but for the 
EDF set 13 (upper panels) and the IBM set 15 (lower panels) in Table~\ref{Tab:NME}. These NME sets are characterized
by relatively low ratios $M_{N,i}/M_{\nu,i}$, and in comparison with those in Fig.~\ref{Fig_06}
provide weaker (stronger) constraints on $m_N$ ($m_\nu$), reflected by the change of scale in the two
coordinates. In the upper panels, the band slopes are very similar to each other, leading to 
an almost complete degeneracy of the two mechanisms. In the lower panels, the degeneracy is partly 
broken, and some limiting cases can be excluded in the $2\sigma$ combination.

The results shown in Fig.~\ref{Fig_06} and~\ref{Fig_07} show that multi-isotope searches 
for $0\nu\beta\beta$ decay with ton-scale experiments have the potential to 
 statistically discriminate two non-interfering mechanisms (the exchange of light and heavy neutrinos),
provided that the corresponding NME are relatively well known and have rather different ratios,
in at least a couple of isotopes. For very similar NME ratios, the mechanisms are instead degenerate.
The future will tell us which conditions are met
by more accurate NME calculations.

\subsection{Analysis with different (true and test) NME sets}

At present, one cannot decide which NME set in Table~\ref{Tab:NME} is the closest
to the one occurring in nature for $0\nu\beta\beta$ decay  (if any). 
This uncertainty may be accounted for by considering different NME sets for the true signals 
$\bar S_i$ and the test signals $S_i$, see e.g. \cite{Faessler:2011rv,Pompa:2023jxc}. In this approach, 
the best fits to the Majorana mass parameters $(m_\nu,\,m_N)$ will generally deviate from the true points in Eq.~(\ref{seed-m})
and the minimum $\chi^2$ will be nonzero.%
\footnote{The value of $\chi^2_\mathrm{min}$ should be referred  
to one degree of freedom, corresponding to three experimental data minus two free mass parameters.}
 Values $\chi^2_\mathrm{min}\gg 1$ would signal that the test NME
provide three allowed bands allowed for (Xe, Ge, Mo) that do not intersect in the same mass parameter region, at least 
for physical values $m^2_{\nu,N}\geq 0$.  
However, values $\chi^2_\mathrm{min}\sim 1$ cannot exclude a priori unfortunate cases, 
where large deviations in the reconstructed parameters are present anyway. 

\begin{figure}[t!]
\begin{minipage}[c]{0.97\textwidth}
\includegraphics[width=0.97\textwidth]{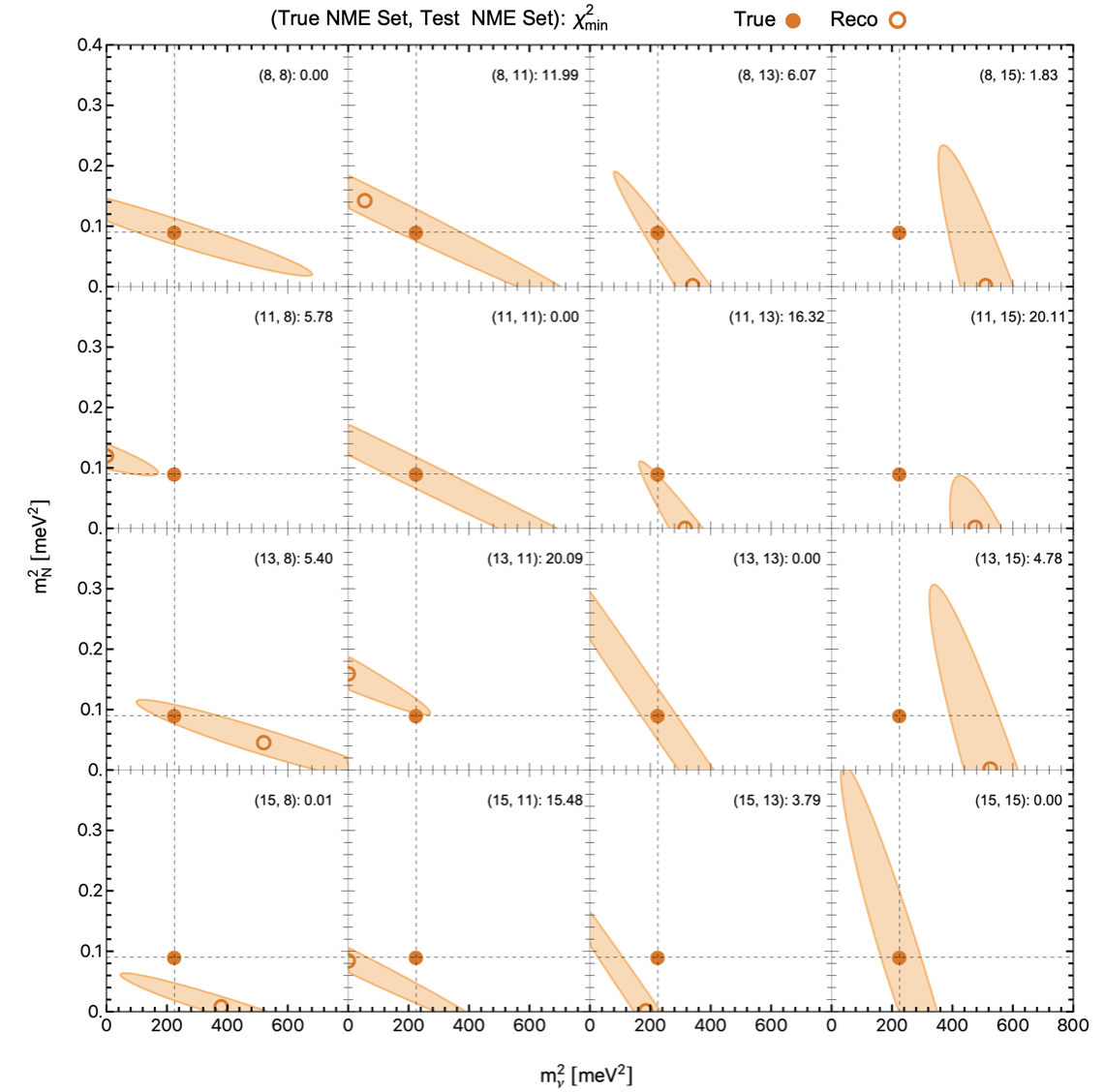}
\vspace*{-2mm}
\caption{\label{Fig_08}
\footnotesize Combined fit to prospective data from nEXO, LEGEND and CUPID, 
assuming the case with both light and heavy neutrino exchange in Eq.~(\ref{seed-m}): $(m_\nu,\,m_N)=(15,\,0.3)$~meV.
Each panel reports a pair of (true,~test) NME sets, numbered as in Table~\ref{Tab:NME}, and followed
by the $\chi^2_\mathrm{min}$ value.  
The true and reconstructed $(m^2_\nu,\,m^2_N)$ points are marked by solid and hollow
circles (coinciding in the diagonal panels); the latter are surrounded by the $2\sigma$ allowed region.
} \end{minipage}
\end{figure}

To study the spectrum of possible outcomes, we have chosen as representative case the scenario with  
both light and heavy neutrino exchange, corresponding to $(m_\nu,\,m_N)=(15,\,0.3)$~meV 
in Eq.~(\ref{seed-m}).  We have combined prospective data from ton-scale experiments, for 
all possible 16 pairs of true and test NME, chosen among the four sets numbered as 8, 11, 13 and 15 in Table~\ref{Tab:NME}.
The results are shown in the 16 panels of Fig.~\ref{Fig_08}. Each panel is identified by a pair of (true,~test) NME sets, 
followed by the $\chi^2_\mathrm{min}$ value. The true values of the Majorana mass parameters are marked 
by a solid circle (the same in all panels), while the reconstructed best-fit values are marked by hollow 
circles, surrounded by the $2\sigma$ allowed region ($\chi^2-\chi^2_\mathrm{min}=4$). Solid and hollow circles coincide in the diagonal panels, where 
the true and test NME sets coincide, and the previous fit results (as shown in
the right panels of Figs.~\ref{Fig_06} and~\ref{Fig_07}) are recovered.

The twelve off-diagonal panels in Fig.~\ref{Fig_08} show a variety of possibilities. 
Three panels correspond to relatively low values $\chi^2_\mathrm{min}\leq 4$, as obtained for 
the NME pairs (8,~15), (15,~8) and (15,~13). The corresponding outcomes are thus
phenomenologically acceptable, but they show a significant bias in the reconstructed values of
the Majorana mass parameters: the best fit is reached at $m^2_N\simeq 0$ (instead of 
$m^2_N = 0.09$~meV$^2$), and the true parameters are well outside the $2\sigma$ allowed region. 
In particular, for the pair (15,~13) the allowed region interpolates smoothly between the 
limits of only light or heavy neutrino exchanges, but misses the true values for the
mass parameters. For the pair (15,~8) note that, despite the significant
differences between the true and test NME sets,
a very good fit $(\chi^2_\mathrm{min}\simeq 0)$ is accidentally obtained.
Four panels in Fig.~\ref{Fig_08} correspond to moderately high values $4<\chi^2_\mathrm{min}\leq 9$,
as obtained for the NME pairs
(8,~13), (11,~8), (13,~8) and (13,~15),
that provide borderline fits to the prospective data. For the first three of these pairs, the reconstructed
mass parameters are within or very close to the $2\sigma$ allowed region, while
for the latter pair the reconstruction bias is quite strong. 
Finally, five panels correspond to high values $\chi^2_\mathrm{min}> 9$,
as obtained for   the NME pairs
(8,~11), (11,~13), (11,~15), (13,~11) and (15,~11). In such cases,
it can be concluded (barring experimental systematics) that the test NME
sets are unable to provide a reasonable description of the data, independently
of any bias in the reconstructed parameters. In this context, having an extra
constraint (three isotopic data versus two free parameters) is crucial to allow a
test of the NME set \cite{Bilenky:2002ga,Bilenky:2004um}. 
Note also that for the pairs (11,~15) and (13,~11), having the
highest $\chi^2_\mathrm{min}$ values, the allowed ellipse is not centered at the reconstructed best-fit point,
and appears to be squeezed towards one axis of the panel. These features signal that the data  fit
would prefer unphysical values, either $m^2_{\nu}<0$ or $m^2_N<0$, if they were formally allowed. 

In conclusion, the spread of numerical NME values in different nuclear models may be  a significant  source
of confusion in the interpretation of future $0\nu\beta\beta$ signals from ton-scale experiments,
assuming that both light and heavy neutrinos contribute to the decays.
In the worst case, one may reconstruct the underlying mass parameters with significant biases
(possibly missing the true parameters), despite an apparently good fit to the data (i.e., a low $\chi^2_\mathrm{min}$).
On the opposite side, very bad fits to the data might allow to reject the chosen NME set
(possibly leading to unphysical parameters). The latter test becomes possible when there are more 
isotopic data than free parameters. Disentangling the complex
interplay between multi-isotope $0\nu\beta\beta$ searches, decay mechanisms, and assumed NME sets,
will require significant progress, especially in the direction of more accurate and 
converging NME calculations. 

Finally, we remark that if future NME calculations will reach a formal accuracy $<20\%$, then the 
conditions suppressing the interference  between the light and heavy neutrinos in LR models
will need to be 
revisited, since residual interference effects \cite{Halprin:1983ez,Ahmed:2019vum}, neglected herein, may become 
comparable to the size of NME uncertainties.

\section{Illustrative test of a theoretical model}
\label{Sec:Model}

In the phenomenological analysis of current and prospective $0\nu\beta\beta$ data, we have made no specific hypothesis on 
the effective Majorana masses $m_\nu$ and $M_N$, treated as free parameters. However, restrictions on the possible values
of $m_\nu$ come from oscillation data in normal ordering (NO) and inverted ordering (IO) for the light $\nu$ masses $m_k$, 
as a function of three unknowns, that following Eq.~(\ref{mlight}) may be chosen as
the lightest neutrino mass,
\begin{equation}
m_L = \min\left\{m_k\right\}_{k=1,2,3}\ ,
\end{equation}
and two relative Majorana phases $\phi_{2,3}\in [0,\,2\pi]$, with
\begin{equation}
\phi_k = \arg(U^2_{ek})-\arg(U^2_{e1}) \ , 
\end{equation}
see e.g.\ the review in \cite{Agostini:2022zub}. 
Moreover, restrictions on admissible pairs
of values $(m_\nu,\,m_N)$ may arise from theoretical models connecting the light and heavy neutrino sectors, such as those
stemming from the see-saw mechanism 
\cite{Minkowski:1977sc,Gell-Mann:1979vob,Yanagida:1979as,Glashow:1979nm,Mohapatra:1979ia,Schechter:1980gr}.
The parameter space of these models can be probed by current joint bounds on $(m_\nu,\,m_N)$, 
and furthermore by prospective $0\nu\beta\beta$ signals in ton-scale experiments.

Within the seesaw mechanism, the mixing between the active and sterile (heavy) neutrinos 
is typically expected to be $V_{eh}^2\sim O(m_k/M_h)$ and is thus significantly suppressed, e.g., if we take 
 $m_k \lesssim O(0.1)~{\rm eV}$, based on  cosmological limits on the sum of light neutrino masses~\cite{Aghanim:2018eyx},
 and $M_h\gtrsim O(1)$~GeV \cite{Bolton:2019pcu}. Larger values of $V_{eh}$ can be attained in specific low-scale seesaw model, see e.g. 
 \cite{Pilaftsis:1991ug, Kersten:2007vk, He:2009ua, Mitra:2011qr}.
A comprehensive phenomenological investigation of joint limits on $M_h$ and $V_{eh}$ can be found in
 \cite{Bolton:2019pcu} (see also \cite{Mitra:2011qr}).
 
While the couplings of heavy $N_h$ with the left-handed (V-A) charged lepton current are generally small, in LR models their couplings with the right-handed (V+A) current are not suppressed, and can have the same size as the active neutrino mixing matrix elements.
On the other hand, the impact of the (V+A) charged current interactions on  observable $0\nu\beta\beta$ decay rates is subject to another kind of suppression, encapsulated by the factor $(m_{W}/m_{W_R})^4$ [see Eq.~(\ref{mheavy})]. 
For the sake of illustration, we consider a specific LR symmetric model with a double seesaw mechanism, as
discussed in \cite{Patra:2023ltl}. For alternative LR models in the context of $0\nu\beta\beta$ decay, using a type-II seesaw mechanism, see e.g.\    \cite{Tello:2010am,BhupalDev:2013ntw,Ge:2015yqa,Li:2020flq}. 

The concerned model \cite{Patra:2023ltl} is constructed so as to satisfy in Eq.~(\ref{mheavy}) the comprehensive $(M_h,\,V_{eh})$ bounds reported in \cite{Bolton:2019pcu}. The model embeds a relatively simple structure for the heavy neutrino sector, that is governed
by the parameters $m_L$ and $\phi_{2,3}$ plus the heaviest mass 
\begin{equation}
M_H = \max\left\{M_h\right\}_{h=1,2,3}\ ,
\end{equation}
where the phenomenological range $M_h\in [1,\,10^3]$~GeV is assumed \cite{Patra:2023ltl}.  
In particular, for each generation ($k=1,\,2,\,3$) and for both NO and IO,
the light and heavy neutrino masses turn out to be inversely proportional, 
\begin{equation}
\frac{m_k}{m_L}=\frac{M_H}{M_k}\ ,
\end{equation}
while the mixing matrix elements are related by
\begin{equation}
V_{ek}= iU^*_{ek}\ .
\end{equation}

Thus, the Majorana mass parameters [Eqs.~(\ref{mlight}) 
and (\ref{mheavy})] admitted by the model satisfy the relation
\begin{equation}
\label{rays}
m_N = \frac{m_e\,m_p}{m_L\,M_H} \left(\frac{m_W}{m_{W_R}} \right)^4 m_\nu\ ,
\end{equation}
corresponding to a bundle of rays in the planes charted by $(m_\nu,\,m_N)$ or
$(m^2_\nu,\,m^2_N)$.

We take the light $\nu$ oscillation parameters at their best-fit points
from \cite{Capozzi:2021fjo}; the effects of the associated uncertainties 
are minor. We also neglect the effects of 
further, very heavy neutrino states associated to the double see-saw structure
of the model \cite{Patra:2023ltl}. The model space is 
then spanned by five free parameters, for both NO and IO:
\begin{equation}
m_{\nu,N}=m_{\nu,N}(m_L,\,M_H,\,m_{W_R},\phi_2,\,\phi_3)\ . 
\end{equation}
We examine a slice of this parameter space by fixing $m_{W_R}=5.5$~TeV
(the lowest limit considered in \cite{Patra:2023ltl}) and 
the heaviest neutrino mass at an intermediate model value, $M_H=300$~GeV. 
The remaining three parameters are randomly sampled in the  
intervals $\phi_{2,3}\in [0,\,2\pi]$ and $m_L\in [0,\,100]$~meV, where
the adopted upper limit on $m_L$ corresponds to $\sum_k m_k \simeq 300$~meV, in
the ballpark of conservative upper bounds  on light neutrino masses from cosmology
\cite{Capozzi:2021fjo}. Cases leading to heavy mass(es) $M_h<1$~GeV are discarded.
The generated model values for $(m^2_\nu,\,m^2_N)$ 
can be compared with the corresponding values allowed by either current Xe+Ge+Te data, or 
by future signals in Xe+Ge+Mo ton-scale experiments. For simplicity, we consider 
only the IBM NME set 15 in Table~\ref{Tab:NME}, with the
associated allowed regions taken from Fig.~\ref{Fig_04} (lower right panel)
and Fig.~\ref{Fig_07} (lower right panel).

\begin{figure}[t!]
\begin{minipage}[c]{0.9\textwidth}
\includegraphics[width=0.98\textwidth]{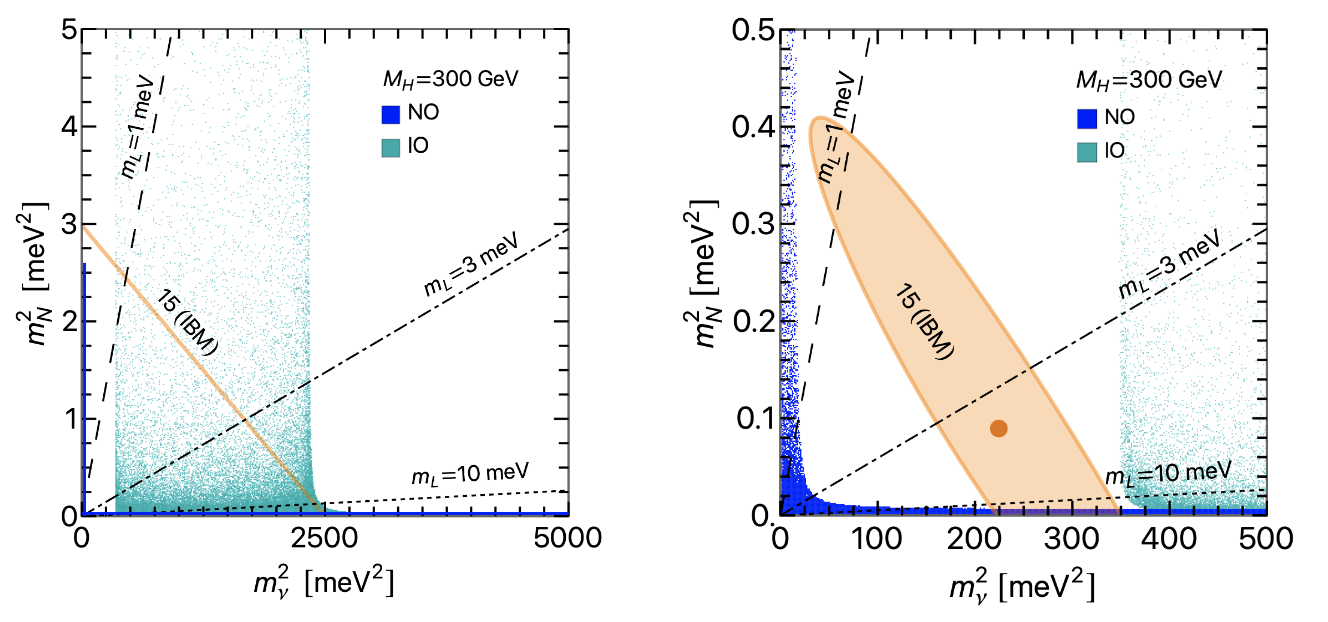}
\vspace*{-2mm}
\caption{\label{Fig_09}
\footnotesize Comparison of specific realizations of the model in \cite{Patra:2023ltl}
with current constraints (left panel) and future signals (right panel) at fixed NME (IBM set 15). 
The blue and cyan points refer to NO and IO, respectively. A few rays at constant $m_L$
are shown. The value of $M_H$ is set at 300 GeV. See the text for details. 
} \end{minipage}
\end{figure}

Figure~\ref{Fig_09} shows the results of these specific model predictions versus current data
(left panel) and
prospective signals (right panel), in the $(m^2_\nu,\,m^2_N)$ plane; note the different scales. Three representative rays [see Eq.~(\ref{rays})] are drawn for $m_L=1,\,3$~and 10~meV. 
In both panels, the sample points for NO and IO are colored 
in blue and cyan, respectively. The NO points are all very close to the axes, so much to
be graphically unresolved in the left panel. The IO points fill a vertical stripe, that bends
towards the right for vanishing $m_N$. In this limit, one recovers 
the well-known fact that $m_\nu$ is bounded from below for IO, but not for NO
\cite{Agostini:2022zub}. 

In the left panel of Fig.~\ref{Fig_09}, points above the orange line are disfavored by
current Xe+Ge+Te data at $\geq 2\sigma$ level; such points include not only NO cases
with $m_N\simeq 0$ and relatively large $m_\nu$, but also interesting IO cases with 
comparable $0\nu\beta\beta$ contributions from light and heavy neutrinos.
Since all ray directions are allowed below the orange line, no significant $0\nu\beta\beta$
constraint can be placed on $m_L$.

In the right panel of Fig.~\ref{Fig_09}, the axis scales (and the model point coordinates) 
are zoomed in by a factor of ten. The elliptic orange region is allowed at $2\sigma$ by prospective Xe+Ge+Mo data. 
In this example, future results tend to
disfavor the IO scenario, not only in the limit of light neutrino exchange (i.e, of vanishing $m_N$), but also
for sizeable contributions of heavy neutrinos to the $0\nu\beta\beta$ signals.  They 
also disfavor NO cases with small $m_\nu$, while allowing NO cases with vanishing $m_N$
and $m^2_{\nu}\simeq 220$--350 meV$^2$. The probability distribution along the ellipse
major axis translates into a probability distribution for $m_L$, 
disfavoring rays with vanishing $m_L$.

Although the above results refer to a specific model for fixed or selected parameters,
compared with representative phenomenological results at fixed NME, they illustrate how 
current and future $0\nu\beta\beta$ searches can probe heavy neutrino physics in theoretically
interesting regions, that are not pre-empted by high-energy constraints.

\section{Summary and conclusions}
\label{Sec:End}

In this work, we have revisited the phenomenology of $0\nu\beta\beta$ decay  
mediated by non-interfering exchange of light $(\nu)$ and heavy ($N$) Majorana neutrinos, using current (Xe, Ge, Te)
data from KamLAND-Zen, EXO, GERDA, MAJORANA, and CUORE, as well as 
prospective (Xe, Ge, Mo) signals in the ton-scale nEXO, LEGEND, and CUPID projects,
within \textcolor{black}{a state-of-the art statistical analysis.}
We have highlighted the interpretation of the results in terms of recent NME sets 
computed in different nuclear models (SM, QRPA, EDF, and IBM). 
In Sec.~\ref{Sec:NME}, we have reported available NME values for light and heavy neutrino mechanisms, and discussed the
corresponding NME ratios, that are
crucially connected to the (non)degeneracy of the mechanisms. 
In Sec.~\ref{Sec:Current} we have derived 
detailed upper bounds on the effective Majorana mass parameters $m_\nu$ and $m_N$, using up-to-date experimental results.
We have shown the usefulness of the plane ($m^2_\nu,\,m^2_N$) to report joint bounds and 
to understand the role of NME ratios. 
In Sec.~\ref{Sec:Future}, we have considered representative cases leading to prospective 
$>3\sigma$ signals in ton-scale projects, for a nominal exposure of 10 ton years. Allowed regions
in the ($m^2_\nu,\,m^2_N$) parameters have been derived and discussed, with emphasis on (non)degenerate 
results. Effects of nuclear model uncertainties have been illustrated by allowing different (true and test) sets of NME values. 
Cases leading to  (un)biased or (un)physical $m_\nu$ and $m_N$ parameters have been discussed. 
The currently large spread of NME values and ratios allows a wide spectrum of outcomes, ranging from reasonable
reconstructions of the Majorana masses to more pessimistic (biased or degenerate) results, that
will hopefully be restricted in the future, while the nuclear models will be improved and benchmarked.  
Finally, in Sec.~\ref{Sec:End}, we have considered a 
specific theoretical model, based on left-right symmetry and connecting the light and heavy neutrino sectors. 
Examples of constraints on its parameter space have been shown, using both current bounds and prospective $0\nu\beta\beta$ 
signals.    
Our findings provide 
further motivations to pursue  multi-isotope $0\nu\beta\beta$ searches 
at the ton mass scale, and to improve the calculations of NME needed for the interpretation of
$0\nu\beta\beta$ decay data in terms of different underlying mechanisms.


\acknowledgments
  
This work is partly supported by the Italian Ministero dell'Universit\`a e Ricerca (MUR) through
the research grant number 2017W4HA7S ``NAT-NET: Neutrino and Astroparticle Theory Network'' under the program PRIN 2017,
and by the Istituto Nazionale di Fisica 
Nucleare (INFN) through the ``Theoretical Astroparticle Physics''  (TAsP) project. 
\textcolor{black}{We are grateful to S.\ Petcov
for reading the manuscript and for useful comments.}  

\vspace*{+8mm}

\end{document}